\documentstyle[prl,aps,multicol,epsf]{revtex}
\input psfig
\def \be{\begin{equation}}
\def \ee{\end{equation}}
\def \r{{\bf r}}
\def \k{{\bf k}}
\def \o{\omega}
\def \p{{\bf p}}

\def \q{{\bf q}}
\def \e{\epsilon}

\def \s1{\sigma_1}
\def \K{G^{(K)}}
\def \R{G^{(R)}}
\def \A{G^{(A)}}
\def \t{\tau_{imp}}
\def \x{{\bf x}}
\begin{document}

\bibliographystyle{simpl1}

\title{Influence of interaction on weak localization.}
\date{\today}
\author{Maxim Vavilov and Vinay Ambegaokar}
\address{Laboratory of Atomic and Solid State Physics,
Cornell University, Ithaca NY 14853, USA}
\maketitle
\begin{abstract}
We discuss the influence of the electromagnetic environment and the
electron--electron interaction on the weak localization correction to the
conductivity of a disordered metal. The theory of this phenomenon for
sufficiently high temperature, where the quantum nature of the interaction
of electrons with the electromagnetic field can be disregarded, has been
understood for some time.  We consider the first order quantum
correction to this semiclassical description and work out the temperature
range in which this correction is small. No external low frequency
cut--off is needed in our calculation. We conclude that in the whole  
region of temperature where the weak localization correction is
much smaller than the Drude conductivity the classical treatment of the
interaction is valid.
\end{abstract}
\draft


\section{Introduction}
Experimental and theoretical studies of  weak localization have given
considerable insight into the physics of small disordered
conductors.  A review of this research can be found in \cite{AAKL}.
Since interference between time-reversed electron trajectories\cite{AAKL,CS} 
is
the root cause of weak localization, its strength depends on phase coherence 
between such paths.  Dephasing can be due to extrinsic causes, such as applied
 magnetic fields, or to intrinsic mechanisms that remove phase information, 
for example scattering against localized spins, or the electron-phonon and 
electron-electron interactions,  the last being the more important at low 
temperatures.

In this paper we reconsider the effect of electron-electron interactions on 
weak localization.  Our aim is to understand quantum corrections to a 
theory \cite{AAK} in which the influence of these interactions is modeled 
as a fluctuating classical electromagnetic field.  The subject is topical 
because of experimental work \cite{MJW} suggesting  that the dephasing rate 
saturates  to a finite value as the temperature approaches zero, whereas  
ref. \cite{AAK}  predicts the rate to vanish in this limit.  Now, one 
certainly expects quantum corrections to the last mentioned theory.  
The relevant question, which we address here, is whether the corrections 
are important in the region of weak localization. Our conclusion, in brief, 
is that they are not.

It is worth comparing our approach to quantum effects with a recent preprint 
\cite{AAG} in which very complete calculations of the quantum mechanical 
correction to the Drude conductivity of a weakly disordered metal are 
carried out to second order in the screened electron-electron interaction.  
In the absence of dephasing, the weak localization correction for narrow 
wires contains an infrared divergence.  This divergence cannot be cured in 
finite order perturbation theory.  For this reason, a magnetic field is 
posited in ref. \cite{AAG}, providing a low frequency cut-off .  The cut-off 
dependence of the correction to weak localization is found to be exactly as 
expected from  \cite{AAK}.  Although we find this argument completely 
convincing, it is nonetheless true that the results of ref. \cite{AAG} 
are, strictly speaking, only valid when the electron-electron interaction
produces a small correction to the effect due to some other mechanism.  [See,
however, Section VI, below.]  Now, the calculation 
of \cite{AAK} has the considerable merit of treating classical fluctuations 
exactly, thereby being free of low frequency divergences.  By building on 
this calculation, we are able to test its validity in the absence of 
extrinsic influences.

According to paper \cite{AAG}, the first order diagrams in the interaction can
be divided onto two parts: 
$\Delta\sigma^{(1)}_{wl}=\Delta\sigma_{deph}+\Delta\sigma_{cwl}$,
$\Delta\sigma_{deph}$ is called the dephasing term and the other 
$\Delta\sigma_{cwl}$
is the interaction correction to weak localization. Although they both 
originate from the same type of diagrams they have different dependences on
the parameters of the system. For example, it was shown in \cite{AAG}, that
for one dimensional wires
\be
\label{b2}
\begin{array}{c}
\displaystyle
\Delta\sigma_{deph}=\frac{e^2\sqrt{D\tau_H}}{\pi\hbar}D^{1/2}\tau_H^{3/2}
\frac{e^2T}{4\hbar^2\sigma_1}
\\
\displaystyle
\Delta\sigma_{cwl}=-3\zeta(3/2)\frac{e^2\sqrt{D\tau_H}}{2\pi\hbar}
\frac{e^2}{2\pi\hbar\sigma_1}\sqrt{\frac{\hbar D}{2\pi T}}.
\end{array}
\ee
Here $\tau_H$ is the dephasing time due to an external magnetic field, which
appears 
in the calculations of \cite{AAG} as a low frequency cutoff. 
We see that the most singular term is the dephasing correction which is 
proportional to $\tau_H^2$, while $\Delta\sigma_{cwl}\sim \sqrt{\tau_H}$.
We consider here only the dephasing term of the weak localization correction
to conductivity. 

The physical meaning of both of these terms was explained in \cite{AAG}.
We would like to note only that the dephasing term is the result 
of the interference of two time reversed paths (see below), while the 
interaction correction term is the result of 
interference for more complicated trajectories.

It is appropriate here to mention some other recent work.  
In e--print \cite{VA1} by us we suggested that the saturation in 
dephasing rate 
\cite{MJW} can be explained in the framework of the Caldeira--Leggett model
\cite{CL}. This paper is simply wrong, because it treats the phase as a single 
particle and loses the physics associated with the exclusion principle.  
The present work is the promised revision of ref. \cite{VA1}.
We also remark that Golubev and 
Zaikin \cite{GZ} have a calculation which claims to treat interaction effects 
to all orders in perturbation theory and obtains a finite dephasing rate from 
the electron-electron interaction at zero temperature.  We believe that this 
surprising result is due to uncontrolled approximations, and comment on it is 
in Appendix C of this paper.  

In outline, the plan of this paper is as follows.  In the next section we 
introduce a model in which the dephasing environment is modeled by a set 
of harmonic oscillators, in the manner of Feynman and Vernon \cite{FV} and 
Leggett and Caldeira \cite{CL}, which are coupled to the charge density of 
an electron gas in a random potential.  By a suitable choice of the spectrum 
of oscillators, we obtain after integrating out the oscillators the influence 
functional corresponding to the diffusively screened electron interaction.  
Up to this point our calculation is identical to that in refs. \cite{AAG} 
and \cite{GZ}.   We deviate from previous work by now separating the 
influence functional into a classical part and a remainder.  In section 
3 we show that the effect of the classical fluctuations on the back 
scattering (or weak localization) correction to the Drude conductivity can be 
written as a path integral, thereby exactly reproducing the result obtained 
in ref. \cite{AAK}.  The improvement over that work is that we have an 
explicit expression for the remainder.  At this point it is convenient to 
specialize the model to the case of Nyquist noise which yields an simpler 
path integral, and allows further calculations to be done analytically.  
This model is used in Section 4 to explicitly calculate the quantum 
corrections to second order.  In Section 5 we give semiquantitative arguments 
for how these calculations are modified when the more realistic spectrum 
corresponding to the screened electron-electron interaction is used, and 
infer the structure of the corresponding quantum corrections up to numerical 
coefficients.  The final section contains a discussion of the results and 
conclusions.

\section{Influence Functional Method for a Disordered Interacting 
Electron Gas }

As the starting point for our calculations we use a variant of the 
Feynman-Vernon \cite{FV},  Leggett-Caldeira \cite{CL} method in which 
a dissipative environment is described by a set of harmonic degrees of 
freedom.  In this way we are able to construct a formalism general 
enough to accommodate different models for the electron-electron interaction.

Consider a closed system described by a  Hamiltonian consisting of three 
parts
\be
\label{1.1}
{\cal H}(t)= {\cal H}_{0}(t) +{\cal H}_{env}(t) + {\cal H}_{int}(t).
\ee
Here the first term describes a disordered free electron subsystem, 
\be
\label{1.2}
{\cal H}_{0}(t)=
\int d\r\psi^+(t,\r)\left(-i\frac{\partial}{\partial t}-
\frac{\nabla^2}{2m}-\mu+U(r)\right)\psi(t,\r),
\ee
where $\mu$ is chemical potential and $U(\r)$ is a random impurity potential. 
 The next term is the Hamiltonian for harmonic electromagnetic modes,
\be
\label{1.3}
{\cal H}_{env}(t)=\sum_{\nu} \o_{\nu}\left(a^+_{\nu} a_{\nu}+
\frac{1}{2}\right),
\ee
where $\nu$ labels different modes of the electric field, which  
correspond to spatial wave functions $\phi_{\nu}(\r)$ which can be 
chosen to be real.
The interaction between the field and the electron system is
\be
\label{1.4}
{\cal H}_{int}(t)=\int d\r\psi^+(t,\r)v(t,\r)\psi(t,\r)=
\int d\r v(t,\r)q(t,\r),
\ee
where $q(t,\r)=\psi^+(t,\r)\psi(t,\r)$ and
\be
\label{1.5}
v(t,\r)=\sum_{\nu}\frac{e}{\sqrt{2M_{\nu}\o_{\nu}}}
\left(\phi_{\nu}(\r)a^+_{\nu}(t)+\phi^*_{\nu}(\r)a_{\nu}(t)\right)
\ee
is an operator in the space of the electric field quantum states.
Above, the Fermion and Boson field operators, $\psi(t,\r)$ and 
$a_{\nu}(t)$, are in the Heisenberg
representation.

The electron Green's function is defined according to
\be
\label{1.6}
G(t,\r,t',\r')=\cases{\displaystyle 
-i\langle \psi(t,\r)\psi^+(t',\r') \rangle, & if $t > _K t'$,\cr
\displaystyle i\langle \psi^+(t',\r')\psi(t,\r)\rangle,
& if $ t<_Kt'$.\cr}
\ee
Here the average $\langle \dots \rangle$ is understood as a trace over the 
quantum state of the whole system, taken with the density matrix $\rho$ 
of the system:
\be
\label{1.7}
\langle \dots \rangle =\frac{{\rm Tr}\left(
\rho T_K \left( \dots \exp(iS[\psi^+,\psi,V])\right)\right)}
{{\rm Tr}\left(
\rho T_K \left(\exp(iS[\psi^+,\psi,V])\right)\right)}, 
\ee 
The action introduced in Eq.(\ref{1.7}) has the form:
\be
\label{1.8}
S[\psi^+,\psi,V]= - \int_K dt \left( {\cal H}_{0}+{\cal H}_{env}+
{\cal H}_{int}
\right).
\ee
In the above formulas $K$ refers to  the Keldysh contour, which 
runs from $-\infty$ to $+\infty$ and then backward to $-\infty$, 
( see, e.g.
\cite{RS} ), and the subscript $K$ refers to ordering along this contour. 
With no loss of generality we may take the $t=-\infty$ initial thermal state 
of the system to be a product of the noninteracting electron density matrix 
$\rho_{el}$ and the environment density matrix $\rho_{env}$, where 
$\rho_{el}$ obeys  Fermi-Dirac statistics and $\rho_{env}$ obeys Bose-Einstein 
statistics

Since the electric field is described by a set of Harmonic oscillators, the 
trace over its quantum states can be worked out exactly.
This yields an influence functional [cf. ref\cite{FV}] for the 
electrons:
\be
\label{1.9}
F[q,q']=\exp(-\Phi[q,q']),
\ee
with
\be
\label{1.10}
\begin{array}{c}
\displaystyle
\Phi[q,q']=\frac{1}{2}\int_{-\infty}^{+\infty}ds\int_{-\infty}^{+\infty} 
du\int d\r_1\int d\r_2 
\\
\displaystyle
\left(
(q(s,\r_1)-q'(s,\r_1))
{\cal K}_1(s-u,\r_1,\r_2)(q(u,\r_2)-q'(u,\r_2))+ \right. \\
\displaystyle
\left.
(q(s,\r_1)-q'(s,\r_1))
{\cal K}_2(s-u,\r_1,\r_2)(q(u,\r_2)+q'(u,\r_2))\right)
\end{array}
\ee
where we have introduced the notations
\be
\label{1.11}
{\cal K}_1(s-u,\r_1,\r_2)=\frac{1}{2}\sum_{\nu}
\frac{e^2}{{2M_{\nu}\o_{\nu}}}\coth\frac{\o_{\nu}}{2T}
\cos(\o_{\nu}(s-u))\phi_{\nu}(\r_1)\phi_{\nu}(\r_2)
\ee
and
\be
\label{1.12}
{\cal K}_2(s-u,\r_1,\r_2)=-i\Theta(s-u)\sum_{\nu}
\frac{e^2}{{2M_{\nu}\o_{\nu}}}\sin(\o_{\nu}(s-u))
\phi_{\nu}(\r_1)\phi_{\nu}(\r_2).
\ee
Above $q(s,\r)=\psi^+(s,\r)\psi(s,\r)$ is the electron density taken at 
the forward part of the Keldysh contour, and 
$q'(s,\r)=\psi^+(s,\r)\psi(s,\r)$
is taken at the backward part of the contour.

Note that a result of the very same form can be obtained for the Coulomb 
interaction
\be
\label{1.12a}
{\cal H}_{int}(t)=\int\int d\r d\r' q(t,\r)V(\r-\r')q(t,\r'),
\ee
where $V(\r)=e^2/|\r|$. The usual way to achieve this can be found in, e.g., 
\cite{ZS}. One performs the Hubbard--Stratanovich transformation to 
decouple the density operators $q(t,\r')$ in Eq.(\ref{1.12a}), introducing 
a fluctuating  electric field. Then the effective action is expanded to 
second order in the fluctuating fields,  and the screened random phase 
approximation is used for 
the electronic polarization.   Finally one integrates over the 
Hubbard--Stratanovich variables to obtain
an effective action for electrons in the form of Eq.(\ref{1.9}), with 
the specific forms of ${\cal K}_1$ and ${\cal K}_2$ given by the choice 
of mode density in Eq.(\ref{1.21}) below.

The influence functional Eq.(\ref{1.10}) is very similar to that obtained 
for a quantum particle coupled to the environment of 
harmonic oscillators. Here, however,
$q(s,\r)$ is not a particle coordinate but  fermion density operator. 
Expanding the exponent of the influence functional $F[q,q']$ we can 
reproduce 
the Keldysh diagram technique for electrons coupled to a Bose field with 
a propagator, which can be expressed in terms of ${\cal K}_1$ 
and ${\cal K}_2$.
One finds that the Keldysh component of the Bose field is ${\cal K}_1$, 
and the retarded component is ${\cal K}_2$. 
It is convenient to use the electronic Keldysh Green's function in the 
`rotated' form \cite{RS}
\be
\label{1.13}
\hat G (t,\r,t',\r')=\left(
\begin{array}{cc}
G ^{(R)}(t,\r,t',\r') & G ^{(K)}(t,\r,t',\r')\\
0& G ^{(A)}(t,\r,t',\r')
\end{array}
\right).
\ee  
One can check that in this representation the vertices corresponding to the
the coupling of electrons  via ${\cal K}_1$ and the vertex at time $s$ 
of the electron coupling via ${\cal K}_2$ in Eq.(\ref{1.10}) 
are proportional to the unit matrix in the Keldysh space, whereas
the vertex corresponding to the electron scattering by the field 
${\cal K}_2(s-u,\r_1,\r_2)$ at time $u$ 
is proportional to the first Pauli matrix $\tau_x$, because of the plus sign 
between $q$ and $q'$ at this vertex in Eq. (\ref{1.10}).

The structure achieved thus far is formally identical to the starting points 
of refs. \cite {AAG} and \cite {GZ}.   At this stage, we make a new
departure by 
explicitly distinguishing between the classical and quantum effects of the 
`environment. Note that at high temperature ${\cal K}_1$ contains the huge 
factor $\coth\o/2T\approx 2T/\o$, so that the second term ${\cal K}_2$ is 
small by comparison.  This fact allows us to represent the effective action 
$F[q,q']$ as a product of classical $F_c[q,q']$ and quantum $F_q[q,q']$ 
parts.
\be
\label{1.14}
F[q,q']=F_c[q,q']F_q[q,q'].
\ee

In $F_q$ we include all of $K_2$  from Eq.(\ref{1.9}) and the part 
of $K_1$ chosen as
\be
\label{1.18}
{\cal K}_{1q}(s-u,\r_1,\r_2)=\frac{1}{2}\sum_{\nu}
\frac{e^2}{2M_\nu\o_\nu}
\left(\coth\frac{\o_{\nu}}{2T}-\frac{2T}{\o_{\nu}}\right)
\cos(\o_{\nu}(s-u))\phi_{\nu}(\r_1)\phi_{\nu}(\r_2).
\ee

The remainder of $K_1$, namely Eq.(\ref{1.11}) with $\coth\o/2T$ replaced 
by $2T/\o$  is included in $F_c$.  Our strategy will be to treat $F_c$ 
exactly, and $F_q$ as a  perturbation.

We now introduce a new  Hubbard--Stratanovich transformation to decrease a 
power of the fermion operators in the exponent of the classical part of the 
influence functional and obtain
\be
\label{1.15}
F_c[q,q']=\int {\cal D}{\bf w}e^{-\sum_\nu w^2_{\nu}}
e^{iS_c[q,q',{\bf w}]},
\ee
where
\be
\label{1.16}
S_c[q,q',{\bf w}]=2\int_K dt\int d\r 
\sum_\nu \frac{e}{\sqrt{2M_\nu \o_\nu}}w_{\nu} q(t,\r)
\sqrt{\frac{T}{2\o_\nu}}\phi_{\nu}(\r)\cos(\o_{\nu}t+\varphi_{\nu})
=\int_K dt\int d\r V(t,\r)q(t,\r),
\ee
\be
\label{1.16a}
V(t,\r)=2e\sum_{\nu}\frac{w_{\nu}}{\sqrt{2M_\nu\o_\nu}}
\sqrt{\frac{T}{2\o_\nu}}
\phi_{\nu}(\r)\cos(\o_{\nu}t+\varphi_{\nu}),
\ee
${\bf w}_{\mu}=(w_{\mu},\phi_{\mu})$ is a two component 
Hubbard--Stratanovich variable for each electromagnetic field mode $\nu$ and
\be
\label{1.17}
\int {\cal D}{\bf w}\dots = \Pi_{\nu}\int\frac{d{\bf w}_{\nu}}{\pi}\dots .
\ee
After these operations, the average in Eq.(\ref{1.7}) reduces to
\be
\label{1.19}
\langle\dots\rangle =\frac{{\rm Tr}\left(
\rho T_K \left( \dots F_c[\psi^+,\psi]F_q[\psi^+,\psi]
\exp(iS_0[\psi^+,\psi])\right)\right)}
{{\rm Tr}\left(
\rho T_K \left(F_c[\psi^+,\psi]F_q[\psi^+,\psi] 
\exp(iS_0[\psi^+,\psi])\right)\right)}
\ee
and $S_0[\psi^+,\psi]= \int_K dt {\cal H}_{0}$.
 
As we have already mentioned, we consider finite order perturbation
theory in $F_q[\psi^+,\psi]$ keeping all orders in $F_c[\psi^+,\psi]$. 
For this purpose it is convenient to introduce classical Green's function
$\hat G_c(t,t',\r,\r')$. It is still defined by Eq.(\ref{1.6}) where the 
average is taken in the sense of Eq.(\ref{1.19}) with 
$F_q[\psi^+,\psi]=1$.
In the Keldysh representation
\be
\label{1.19a}
\hat G_c (t,\r,t',\r')=\left(
\begin{array}{cc}
G_c ^{(R)}(t,\r,t',\r') & G_c ^{(K)}(t,\r,t',\r')\\
0& G_c ^{(A)}(t,\r,t',\r')
\end{array}
\right)
\ee
and 
\be
\begin{array}{c}
\label{1.19b}
\displaystyle
G_c ^{(R)}=\R_0+\R_0 V \R_0 + \dots \\
\displaystyle
\K_c=\K_0+\R_0 V \K_0 + \K_0 V \A_0 +\R_0 V\dots \R_0 V \K_0 V \A_0 
\dots V \A_0+\dots \\
\displaystyle
G_c ^{(A)}=\A_0+\A_0 V \A_0 + \dots ,
\end{array}
\ee
where $\hat G_0$ is the Green's function of electron in metal without 
interaction.

To conclude this section we discuss two physically motivated choices for the 
density of environmental modes. Instead of the summation over mode index 
$\nu$ we integrate over $\o$ and $q$, where $\o$ is a frequency and $q$  
a wave vector, and write
\be
\label{1.20}
\sum_{\nu}\frac{e^2}{2 M_{\nu}\o_{\nu}}{\cal F}(\o_{\nu},q_{\nu}) = 
\int\frac{d\o}{2\pi}\int\frac{dq}{2\pi}J(\o,q){\cal F}(\o,q).
\ee
The choice
\be
\label{1.21}
J(\o,q)=(2\pi)^2\sum_\nu 
\frac{e^2}{2M_\nu 
\o_\nu}\delta(\o-\o_\nu)\delta(q-q_\nu)=\frac{e^2}{\s1}\frac{\o}{q^2},
\ee
corresponds to the low frequency and small momentum 
spectral function of the screened Coulomb interaction in a disordered metal,
with $\s1$   the one dimensional conductivity of the wire.  

Both for its own interest and because it gives an analytically simpler 
structure, we shall also consider the dissipative effect of Nyquist noise 
associated with an external resistor.  We note that the spectral function 
of the 
electric field responsible for the Nyquist noise can be described by 
Eq.(\ref{1.21}) with $q<q_m=2\pi\kappa /L$ with $\s1$ having the dimension
of a 
1D conductivity, $\kappa\ll 1$. This choice of $\kappa$ allows us to take the
limit of a uniform electric field inside the wire, corresponding to the
Nyquist noise at the terminals of the wire produced by external part of
the electric circuit. To see this, calculate the average value 
of the voltage $U=v(L)-v(0)$ at the ends of the wire, where $v(x)$ is 
defined by 
Eq.(\ref{1.5}):
\be
\label{1.22}
\begin{array}{c}
\displaystyle
\langle U_{\o}^2 \rangle=\langle (v(0)-v(L))_{\o}^2 \rangle=
\\
\displaystyle
2e^2\int_0^{q_{m}}  (1-\cos(qL))\frac{\o}{\s1}\frac{1}{q^2}\frac{dq}{2\pi}
\coth\frac{\o}{2T}
\approx e^2 \o R_{eff}\coth\frac{\o}{2T}.
\end{array}
\ee
Here $R_{eff}=\kappa L/ \s1$ is now to be thought of as
determined by 
resistances of the circuit.  When $R_{eff} $ is thus defined we see that
the voltage fluctuations obey the 
fluctuation--dissipation theorem, see \cite{LL9}, appropriate to Nyquist 
noise.

For a wire with resistance $R_w$ connected with resistor $R_0$ one has, 
see \cite{AAK}
\be
\label{1.23}
R_{eff}=\frac{R_w^2 R_0}{(R_0+R_w)^2}.
\ee

\section{Classical result for the weak localization correction to the
conductivity}

In this section we calculate the weak localization correction to the 
conductivity including the classical part of the influence functional, 
defined by Eqs. (\ref{1.1}), exactly but neglecting the quantum part.  
We shall show that the environment now behaves like a classical 
electromagnetic field,  so that the calculation---repeated here for 
completeness---is very similar to the one done in \cite{AAK}.  A small 
difference is that in our representation the fluctuating field is described 
by a scalar potential $V(t,\r)$ instead of a vector potential. 

The current operator is given by
\be
\label{2.1}
\hat {\bf j}(\r)= \frac{ie}{2m}\left(\nabla_{\r}-\nabla_{\r'}\right)_{\r=\r'}
-\frac{e^2{\bf A}(\r)}{m},
\ee
where ${\bf A}(\r)$ is the vector potential corresponding to an external 
field which produces an average current given by  
\be
\label{2.2}
{\bf j}(\r)=\hat{\bf j}(\r) \int\frac{d\e}{2\pi}\K_1(\e,\r,\r')_{\r=\r'}.
\ee
To calculate the linear response to the vector potential ${\bf A}(\r)$ 
it is sufficient 
to find the Keldysh component of the electron Green's function  to 
the first order:
\be
\label{2.3}
\begin{array}{c}
\displaystyle
\K_1(\e,\r,\r')=\K (\e,\r,\r')+\int d\r_1\R (\e,\r,\r_1)e{\bf A}(\r_1)
\hat{\bf j}(\r_1)\K (\e-\o_{ext},\r_1,\r')+
\\
\displaystyle
\int d\r_1\K (\e,\r,\r_1)e{\bf A}(\r_1)\hat{\bf j}(\r_1)
\A (\e-\o_{ext},\r_1,\r').
\end{array}
\ee 
We have supposed that the external field oscillates with a finite 
but small frequency $\o_{ext}$, so that $A=cE/i\o_{ext}$.  
Because we are dealing with a 
non-superconducting disordered system, the diamagnetic (i.e. the second)
term of the current operator on the right hand side of Eq.(\ref{2.1}) cancels
a $\o_{ext}^{-1}$ contribution to the conductivity from the first term.
Note that $\hat G (\e,\r,\r')$ is the exact electron Green's function 
of the 
electron--environment system, defined by Eq.(\ref{1.6}).
Since in this section we consider $F_q[\psi^+,\psi]=1$, we can replace 
$\hat G (\e,\r,\r')$ by $\hat G_c (\e,\r,\r')$, defined in Eq.(\ref{1.19a}).

We shall treat the impurity potential $U(\r)$ and electric field $V(t,\r)$ 
defined in Eq.(\ref{1.16a}) as perturbations, keeping all orders in them.
The corresponding electron vertices are proportional to the unit 
matrix in the Keldysh space.   From this observation it follows that
there is only one Keldysh component in every conductivity diagram. 

We consider here the weak localization correction to conductivity which 
is given by a maximally crossed diagram, see Fig.1. 
\begin{figure}
\centerline{\psfig{figure=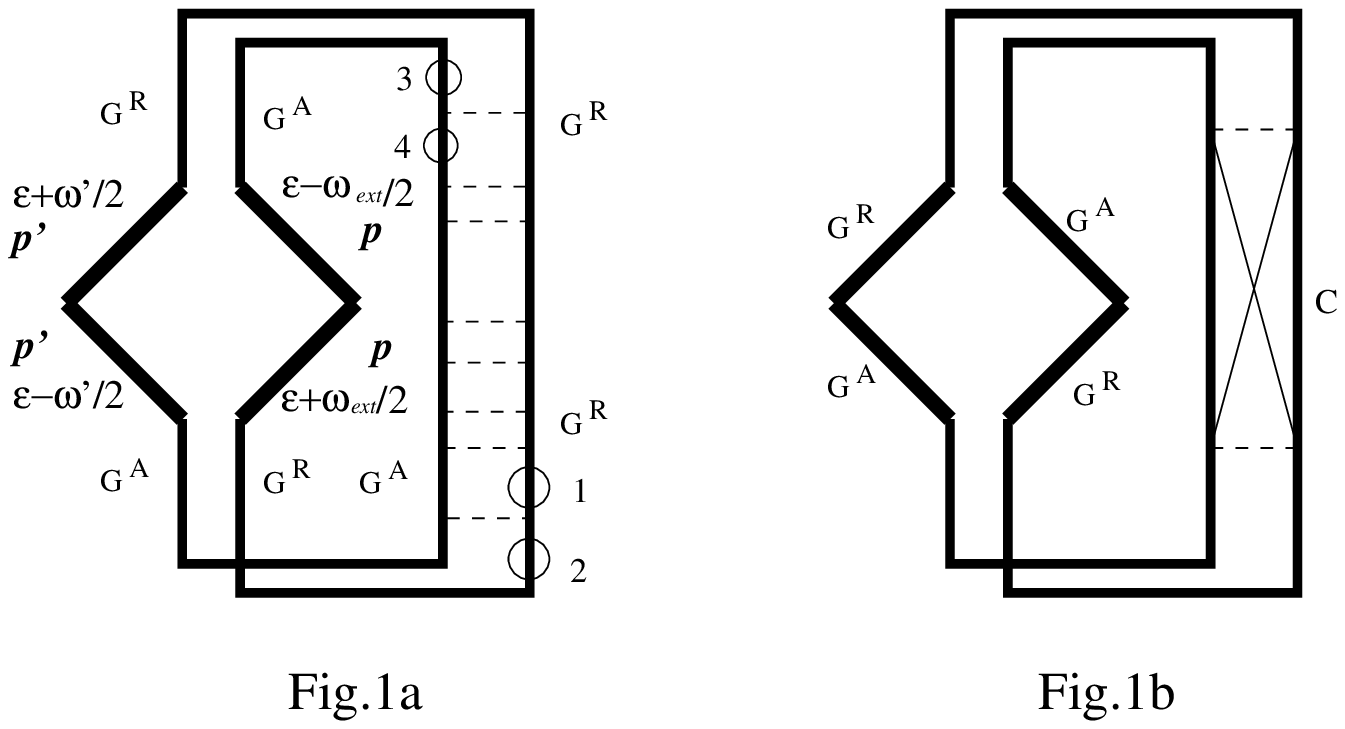,width=8.5cm}}
\caption{
Figure 1a shows maximally crossed diagrams which 
correspond to the weak localization correction to conductivity.
Open circles denote the position for the Keldysh component of the 
electron Green's function. The sum of all possible diagrams is 
represented by the diagram in Fig.1b. The Cooperon $C$ satisfies the equation 
illustrated in Fig.2.
}
\end{figure}
\begin{figure}
\centerline{\psfig{figure=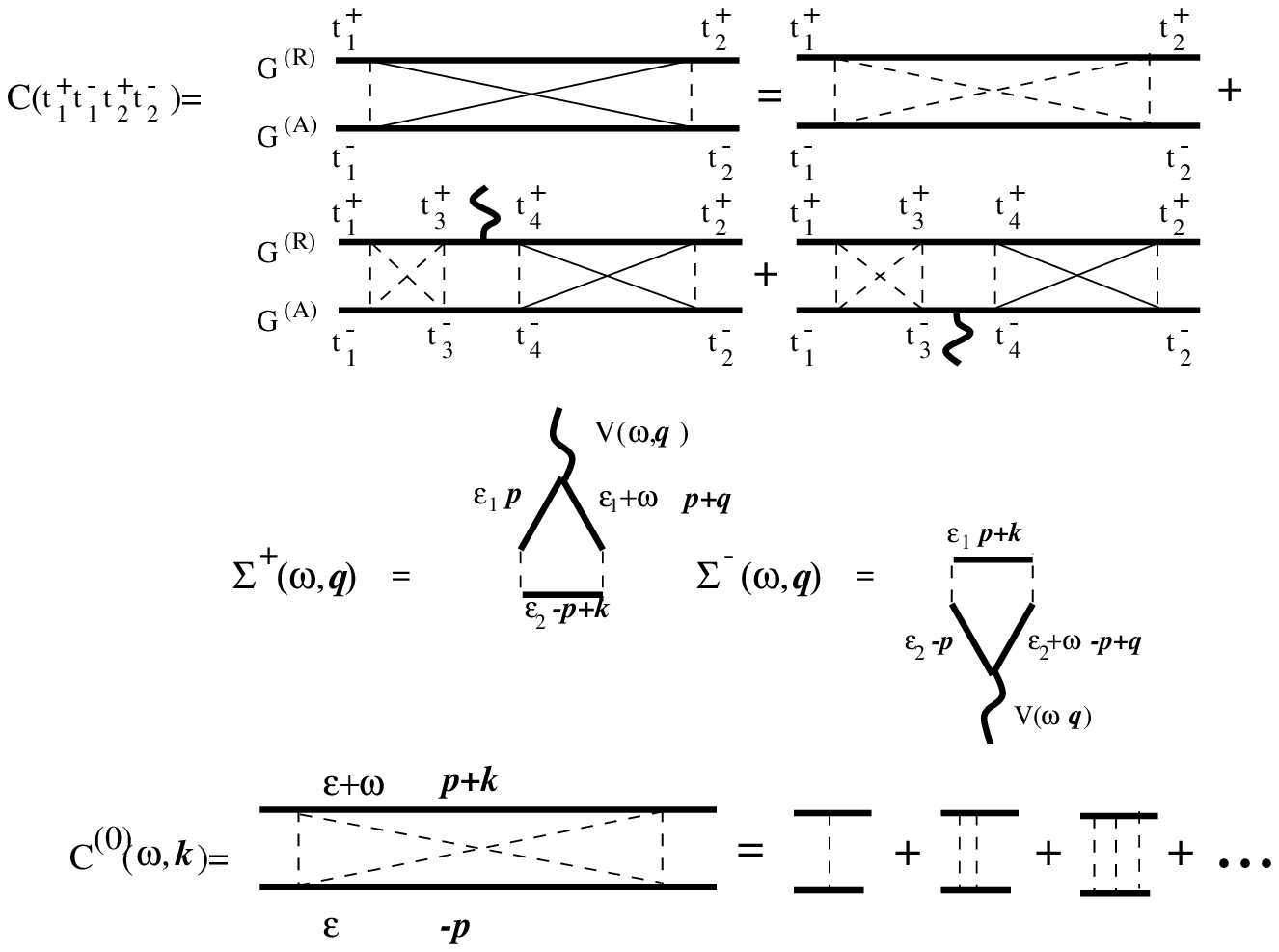,width=8.5cm}}
\caption{
The Dyson equation for the Cooperon in the classical 
electric field $V(t,\r)$.
}
\end{figure}
The first term in Eq.(\ref{2.3}) does not contribute to this correction 
and two other terms do not vanish only if the
Keldysh 
component stands at the places marked by circles in Fig.1. Otherwise we get 
integrals 
\be
\label{2.5}
\int\frac{d{\bf p}}{(2\pi)^d}\A(\e,\p)\A(\e',\p+\k)=
\int\frac{d{\bf p}}{(2\pi)^d}\R(\e,\p)\R(\e',\p+\k)=0. 
\ee 

The Green's functions in Eq.(\ref{1.19b}) 
are also impurity renormalized so that
\be
\label{2.6}
G^{(R,A)}(\e,\p)=\frac{1}{\e-\e(p)\pm i/2\t},
\ee
and $\e(\p)=\p^2/2m-\mu$ is the energy of a free electron with momentum $\p$.
In equilibrium the Keldysh component satisfies the equation
\be
\label{2.7}
\K(\e,\p)=h(\e)\left(\R(\e,\p)-\A(\e,\p)\right),
\ee
where $h(\e)=\tanh \e/2T$. 
 
Using Eq.(\ref{2.7}) we note that the diagram, which contains the Keldysh 
component at the position marked by 1 is canceled by a part of the diagram 
with the Keldysh component at position 2. There is a similar cancellation of 
some terms of the remaining two diagrams. The result can be represented in 
the 
form, shown in Fig.(2). We will keep the leading terms in 
$\o \t$ and $D \k^2 \t$, where $\o$ and $\k$ are the characteristic energy 
and 
momentum changes due to the fluctuating field.

The corresponding analytical expression for the weak localization correction 
to the conductivity is given by
\be
\label{2.8}
\begin{array}{c}
\displaystyle
\Delta\sigma_{wl}(\o')=e^2\int\frac{d\p d\e}{(2\pi)^{d+1}}
\int\frac{d\p' d\e'}{(2\pi)^{d+1}}\p\cdot\p'
{\cal M}(\e,\e',\p,\p')\frac{\left(h(\e+\o_{ext}/2)-h(\e-\o_{ext}/2)\right)}
{\o_{ext}} 
\\
\displaystyle
C(\e'+\o '/2,\e-\o_{ext}/2,\e+\o_{ext}/2,\e'-\o '/2,\p+\p').
\end{array}
\ee
Here 
\be
\label{2.9}
{\cal M}(\e,\e',\p,\p')=\R(\e+\o_{ext}/2,\p)\A(\e-\o_{ext}/2,\p)
\R(\e'+\o'/2,\p')\A(\e'-\o'/2,\p')
\ee
is known as a Hikami box, 
$C(\e_1,\e_2,\e_3,\e_4,\p+\p')$ is the Cooperon, $\o'$ is the current 
frequency. Since the Cooperon is 
singular function of $\p+\p'$, we can use the equality $\p\approx -\p'$ 
everywhere,
except as an argument of the Cooperon.  
Performing the integration over $\p$ and keeping the Cooperon dependence on 
$\p+\p'=\k$, we get 
\be
\label{2.10}
\begin{array}{c}
\displaystyle
\Delta\sigma_{wl}(\o')=-4\pi\nu D e^2\t^2 \int\frac{d\k}{(2\pi)^d}
\int\frac{d\e}{2\pi}
\frac{\left(h(\e+\o_{ext}/2)-h(\e-\o_{ext}/2)\right)}{\o_{ext}}
\\
\displaystyle
\int\frac{d\e'}{2\pi}
C(\e'+\o'/2,\e-\o_{ext}/2,\e+\o_{ext}/2,\e'-\o_{ext}'/2,\k).
\end{array}
\ee

Further calculations are more convenient in the time representation. We 
define the Fourier transform of the Cooperon by the equation:
\be
\label{2.11}
\begin{array}{c}
\displaystyle
C(\e_1,\e_2,\e_3,\e_4,\r,\r')=\int\int\int\int dt_1^+dt_1^-dt_2^+dt_2^-
\exp(i(\e_1t_1^++\e_2t_1^--\e_3t_2^+-\e_4t_2^-)) \times
\\
\displaystyle
C(t_1^+,t_1^-,t_2^+,t_2^-,\r,\r').
\end{array}
\ee 
Not all four time variables of the Cooperon are independent
and we introduce a new notation
\be
\label{2.12}
C(t_1^+,t_1^-,t_2^+,t_2^-,\r,\r')={\cal C}(T,\eta_1,\eta_2,\r,\r')
\delta(t_1^++t_1^--t_2^+-t_2^-), 
\ee
where $2T=t_1^++t_1^-$, and $\eta_{1,2}=t_{1,2}^+-t_{1,2}^-$.
For the present purposes one only needs this function for 
$\eta_1=-\eta_2= \eta$
This can be seen by integration over $\e'$ in Eq.(\ref{2.10}) 
producing the constraint $\delta(\eta_1+\eta_2)$, and allowing one time 
integral to be done.  Then one can also complete the integration over $\e$, 
since the only remaining dependence on $\e$ is in the difference of the 
electron distribution functions $h(\e+\o_{ext}/2)$ 
and $h(\e-\o_{ext}/2)$.  After these steps one reaches the formula
\be
\label{2.18}
\Delta\sigma_{wl}(\o_{ext}=0)=-\frac{4e^2D}{\pi}\int_0^{\infty}d\eta P(\eta),
\ee
where $P(\eta)$, which has the meaning of the probability of returning to 
a starting point after a time $\eta$, is the Cooperon evaluated at 
coinciding space and time points,
\be
\label{2.17}
P(\eta)=2\pi\nu\t^2 {\cal C}(T,+\eta,-\eta,\r,\r)
\ee
To evaluate the Cooperon we can use the following path integral form 
(see Appendix A for more details).
\be
\label{2.13}
{\cal C}(T,\eta_1,\eta_2,\r,\r')
=\frac{1}{2\pi\nu\t^2} \int_{r(\eta_2)=\r'}^{r(\eta_1)=\r}{\cal D}\r(t)
\exp\left(-\int_{\eta_2}^{\eta_1}
d\zeta\left(\frac{\dot\r^2 (\zeta)}{4D}-i\tilde 
V(T,\zeta,\r(\zeta))\right)
\right),
\ee
with
\be
\label{2.14}
\tilde V(T,\zeta,\r(\zeta))=V(T+\zeta /2,\r(\zeta))-V(T-\zeta /2,\r(\zeta)),
\ee 
and the electric field potential $V(t,\r)$ defined by Eq.(\ref{1.16a}).  
In performing the integration over the magnitude and the phase of the 
electric field
according to Eq.(\ref{1.15}), we note that the time variable $T$ can be 
absorbed into the phase $\phi_{\nu}$ of the field. Consequently, the 
correction to the conductivity is independent of $T$.  As the result of 
the integration we get for a wire, in which case $\r$ is one dimensional,
\be
\label{2.15}
{\cal C}(+\eta,-\eta,r,r')=\int_{r(-\eta)=r'}^{r(+\eta)=r}{\cal D}r(t)
\exp\left(-\int_{-\eta}^{+\eta}dt\left(\frac{\dot r^2 (t)}{4D}+
U\left(r, t  \right)
\right) \right),
\ee
where
\be
\label{2.15a}
U(r,t)=\cases{\displaystyle
\frac{e^2TR_{eff}}{ L^2}(r(t)-r(-t))^2, & for the Nyquist noise model,\cr
\displaystyle \frac{2e^2T}{\s1}|r(t)-r(-t)|, & for the electron--electron 
interaction.\cr}
\ee
Here we have used the specific time variables $\eta_1=\eta$ and 
$\eta_2=-\eta$, needed for the weak localization correction to the 
conductivity.
 
At this point it is convenient to introduce new variables for the path 
coordinates. Let us define $R(t)=(r(t)+r(-t))/\sqrt{2}$ and 
$x(t)=(r(t)-r(-t))/\sqrt{2}$. Then we can eliminate the 
integration over negative time. Also this change of variables separates 
$R(t)$ and $x(t)$ in the exponent. The motion described by $R(t)$ is an 
ordinary 
diffusion and a direct integration gives unity for the whole integral.  
The  path integral over $x(t)$ can be done easily for the Nyquist noise 
model, because the motion corresponding to $x(t)$ is  that of a particle in 
a harmonic oscillator which is at the position of minimum potential energy 
at the starting and final points.  Thus, Nyquist noise yields the following  
simple explicit result for the Cooperon
\be
\label{2.16}
\begin{array}{r}
\displaystyle
{\cal C}(\eta,-\eta,\r,\r')=\frac{\sqrt{2}}{4\pi\nu\t^2}\int_{0}^{x(\eta)=
(\r'-\r)/\sqrt{2}}
{\cal D}x(t)\exp\left(-\frac{1}{4D}\int_{0}^{\eta}dt\left(\dot x^2 (t)+
\Omega^2 x^2(t)  \right)\right)=\\
\displaystyle
\frac{1}{2\pi\nu\t^2}
\sqrt{\frac{\Omega}{8\pi D\sinh \Omega \eta}}
\exp\left(-\frac{\Omega (r'-r)^2}{8D}\coth\Omega \eta\right),
\end{array}
\ee
where 
\be
\label{2.16a}
\Omega^2= 16 \frac{ e^2 D T R_{eff}}{ L^2}=16 TT_q
\ee 
and $T_q= DTe^2R_{eff}/L^2$.

We thus have
\be
\label{2.16b}
P(\eta) = \sqrt{\frac{\Omega}{8\pi D\sinh \Omega \eta}},
\ee
from which it follows that in this model
\be
\label{2.16c}
\Delta\sigma_{wl} = -\frac{\sqrt{2}e^2}{\sqrt{\pi ^3}}
I_0\sqrt{\frac{D}{\Omega}},
\ee
where 
\be
\label{2.19}
I_0=\int_0^\infty \frac{dx}{\sqrt{\sinh x}}\approx 3.71.
\ee

So far we have assumed that the interaction with the 
environment is the only source of the dephasing.  If the sample is in a
magnetic field $H$ we have to consider the competition between the 
dephasing produced by the interaction and the magnetic field. The magnetic
field exponentially suppresses the Cooperon ( Eq.(\ref{2.13}) ) as a 
function 
of $\eta_2-\eta_1$. The magnetic characteristic time is 
$\tau_H^{-1}=e^2DH^2a^2/3c^2$, where $a$ is the thickness of the wire, see 
\cite{AAKL}. When both sources of dephasing are present the weak localization 
correction to conductivity is given by
\be
\label{2.20}
\Delta\sigma_{wl}=-\frac{4e^2D}{\pi}\int_0^{\infty}d\eta
P(\eta)e^{-2\eta/\tau_H}.
\ee 

This formula allows us to reproduce in a different way from theirs some 
results of \cite{AAK} for $\Delta\sigma_{wl}$ in two limits.

 For $\tau_H\Omega \ll 1$
we can expand $(\sinh \Omega\eta)^{-1/2}=1/\sqrt{\Omega\eta}-
(\Omega\eta)^{3/2}/12$ and obtain
\be
\label{2.21}
\Delta\sigma_{wl}=-\frac{e^2}{\pi}\sqrt{D\tau_H}
\left(1-\frac{1}{64}(\tau_H\Omega)^2\right).
\ee 
In the opposite limit, $\tau_H\Omega \gg 1$
we expand $\exp(-\eta/\tau_H)\approx 1-\eta/\tau_H$ and integrate over 
$\eta$ to get
\be
\label{2.22}
\Delta\sigma_{wl}=-\frac{\sqrt{2}e^2}{\pi^{3/2}}\sqrt{\frac{D}{\Omega}}
\left(I_0-I_1\frac{2}{\tau_H\Omega}\right),
\ee
where
\be
\label{2.23}
I_1=\int_0^\infty \frac{\zeta} {\sqrt{\sinh \zeta}} d \zeta\approx 5.84.
\ee

In the case of the electron--electron interaction, the functional integral 
has been related to a Schr\"odinger-like differential equation in ref. 
\cite{AAK} with the result
\be
\label{4.5}
\Delta \sigma_{wl}^{ee}
=\frac{e^2}{\pi}\sqrt{D\gamma^{-1}_{ee}}
\frac{1}{\left(\ln{\rm Ai}(1/\tau_H\gamma) \right)^{\prime}},
\ee
where ${\rm Ai}(x)$ is the Airy function and 
\be
\label{2.24}
\gamma_{ee}=\left(\frac{e^2T\sqrt{D}}{\sigma_1} 
\right)^{2/3}.
\ee
$L_{ee}=\sqrt{D/\gamma_{ee}}$ and $L_H=\sqrt{D\tau_H}$.

In weak magnetic field $L_H\gg L_{ee}$ 
\be
\label{2.25}
\Delta \sigma_{wl}^{ee}
=-\frac{2e^2 L_{ee}}{3^{5/6}\Gamma^2(2/3)}.
\ee
where $L_{ee}=\sqrt{D/\gamma_{ee}}$ and $L_H=\sqrt{D\tau_H}$.
In the opposite case we use 
\be
\label{2.26}
{\rm Ai}(x)\sim\frac{1}{2(\pi^2 x)^{1/4}}e^{-2/3 x^{3/2}}
\ee
to get 
\be
\label{2.27}
\Delta \sigma_{wl}^{ee}=-\frac{e^2 L_{H}}{\pi}
\left( 1-\frac{1}{4}(\tau_H\gamma)^{3/2}\right).
\ee
Note, that the result of Eq.(\ref{2.24}) differs from one found in \cite{AAK}
by a factor of 2. One the other hand the expansion Eq.(\ref{2.27}) is
consistent with the result of \cite{AAG}. Also the result of Eq.(\ref{2.16c})
has an extra $2^{-1/4}$ to the numerical factor 
$\Gamma(1/4)/2\pi\Gamma(3/4)$ found in \cite{AAK}
for the weak localization correction to conductivity in the presence of
the Nyquist noise. 

We have thus seen that our 'classical' Action exactly reproduces known high
temperature results for the intrinsic dephasing effect in weak localization 
without the need for an external infrared cut-off.  In the remainder of this 
paper we build
on these calculations to examine quantum corrections. 

\section{Quantum corrections}

Now we consider the contribution of the quantum part of the influence 
functional to the weak localization correction to conductivity. 
In this section we consider dephasing correction to conductivity due to the 
Nyquist noise. In this case the calculations can be done analytically.
The calculations described here are similar to those in ref. 
\cite{AAG}. 
The difference is that ours are free from infrared divergences, because the 
low frequency modes of interaction have already been taken into account 
exactly.

Expanding $F_q[\psi^+,\psi]$, the second term in Eq.(\ref{1.14}), to the 
first order we obtain
\be
\label{3.1}
\begin{array}{c}
\displaystyle
F_q[\psi^+,\psi]\approx 1-\frac{1}{2}
\int_{-\infty}^{+\infty}ds\int_{-\infty}^{+\infty} 
du\int d\r_1\int d\r_2 \\
\displaystyle
\left(
{\cal L}^{(K)}(s-u,\r_1,\r_2)(q(s,\r_1)-q'(s,\r_1))(q(u,\r_2)-q'(u,\r_2))+
\right.\\
\displaystyle
{\cal L}^{(R)}(s-u,\r_1,\r_2)(q(s,\r_1)-q'(s,\r_1))(q(u,\r_2)+q'(u,\r_2))+
\\
\displaystyle
\left.
{\cal L}^{(A)}(s-u,\r_1,\r_2)(q(u,\r_1)-q'(u,\r_1))(q(s,\r_2)+q'(s,\r_2))
\right),
\end{array}
\ee
where Keldysh labels have been assigned according to the definitions
\be
\label{3.2}
\begin{array}{c}
\displaystyle
{\cal L}^{(R)}(t,\r_1,\r_2)=\Theta(t)\sum_{\nu}
e^2\frac{\phi_{\nu}(\r_1)\phi_{\nu}(\r_2)}{4M_{\nu}\o_{\nu}}\exp(-i\o_{\nu}t),
\\
\displaystyle
{\cal L}^{(A)}(t,\r_1,\r_2)=-\Theta(-t)\sum_{\nu}
e^2\frac{\phi_{\nu}(\r_1)\phi_{\nu}(\r_2)}{4M_{\nu}\o_{\nu}}\exp(-i\o_{\nu}t),
\\
\displaystyle
{\cal L}^{(K)}(t,\r_1,\r_2)=\sum_{\nu}
e^2\frac{\phi_{\nu}(\r_1)\phi_{\nu}(\r_2)}{4M_{\nu}\o_{\nu}}
\left(\coth\frac{\o_{\nu}}{2T}-\frac{2T}{\o_{\nu}}\right)
\cos(\o_{\nu}t).
\end{array}
\ee

As we have already mentioned in the previous section, the conductivity is 
determined by the Keldysh component of the electron Green's function ( see 
Eq.(\ref{2.2}) ). The Keldysh component can be represented in terms of the
Green's function defined in Eq.(\ref{1.6}):
\be
\label{3.2a}
\K(t,t',\r,\r')= G(t_f,t'_b,\r,\r')+G(t_b,t'_f,\r,\r')
\ee
where the $f-$index of time argument means that it is taken at the 
forward part of the Keldysh contour, and the $b-$ index for the backward
part, so that $t_f<_K t_b$.
The average in Eq.(\ref{1.6}) is represented by Eq.(\ref{1.19}) with respect to
the classical part of the effective action $F_c[\psi^+,\psi]$, see 
Eq.(\ref{1.15}) and the quantum part in the form of Eq.(\ref{3.2}).
The first term in Eq.(\ref{3.2}), which is unity, provides the expression
for $\K(t,t',\r,\r')$ considered in the previous section and the remaining 
part of Eq.(\ref{3.2}) gives the first order quantum correction. 
Using the Wick's theorem for the electron operators we get after cumbersome 
calculations the quantum correction to the Keldysh component of the Green's 
function 
\be
\label{3.3}
\begin{array}{l}
\displaystyle
\K_{1\times int}(\x,\x')=
\int d\x_1\int d\x_2 
\left(
{\cal L}^{(K)}(\x_1,\x_2)\left(\R_1(\x,\x_1) 
\R_1(\x_1,\x_2)\K_1(\x_2,\x')
+\right.\right.
\\
\displaystyle
\left.
\R_1(\x,\x_1)\K_1(\x_1,\x_2)\A_1(\x_2,\x')+
\K_1(\x,\x_1)\A_1(\x_1,\x_2)\A_1(\x_2,\x')\right)+
\\
\displaystyle
{\cal L}^{(R)}(\x_1,\x_2)\left(\R_1(\x,\x_1)\K_1(\x_1,\x_2)
\K_1(\x_2,\x')+
\R_1(\x,\x_1)\R_1(\x_1,\x_2)\A_1(\x_2,\x')\right. +
\\
\displaystyle
\left.
{\cal L}^{(A)}(\x_1,\x_2)\left(\K_1(\x,\x_1)\K_1(\x_1,\x_2)
\A_1(\x_2,\x')+
\R_1(\x,\x_1)\A_1(\x_1,\x_2)\A_1(\x_2,\x')\right)\right),
\end{array}
\ee
where $\x=(t,\r)$.
Subscript $_1$ means that the Green's functions should be calculated up to
the first order in the external electric field ${\bf A}(\r)$, similarly to
what has been done in Eq.(\ref{2.2}). 
Thus, the Keldysh component is given by Eq.(\ref{2.3}) and 
$G^{(R,A)}_1(t,t',\r,\r')$ are
\be
\label{3.3a}
\begin{array}{c}
\displaystyle
\R_1(t,t',\r,\r')=\R_c(t,t',\r,\r')+\int d\r_1\int dt_1\R_c(t,t',\r,\r_1)
e {\bf A}(t_1,\r_1) {\bf j}(\r_1)\R_c(t_1,t',\r_1,\r'),
\\
\displaystyle
\A_1(t,t',\r,\r')=\A_c(t,t',\r,\r')+\int d\r_1\int dt_1\A_c(t,t',\r,\r_1)
e {\bf A}(t_1,\r_1) {\bf j}(\r_1)\A_c(t_1,t',\r_1,\r')
\end{array}
\ee

Eq.(\ref{3.3}) is in agreement with the expression for the
first order correction to the Keldysh component of the electron Green's
function obtained in the standard technique.  [See \cite{RS} for a
general discussion and \cite{AAG} for the explicit form.]
  
It was shown in paper \cite{AAG} that to second order in the interaction
the weak localization correction to the conductivity can be represented as a
sum of two terms, called the dephasing term and the cross term. The
dephasing term contains the contribution from the electron interaction with
Boson modes having energies smaller than temperature of the system. On the
other hand the cross term has contribution from the whole spectrum, and
corresponds to the electron scattering from the Friedel oscillations.
According to \cite{AAG} [c.f. Eq.(\ref{b2}) above,] 
the interaction correction term has a weaker divergence, 
$\Delta\sigma_{cr} \sim \sqrt{\tau_{\varphi}}$ (where $\tau_\varphi$ is the
decoherence time), than the dephasing term $\Delta\sigma_{deph}\sim
\tau_{\varphi}^2$.  Keeping this in mind, we restrict our 
calculations to this latter term, which is expected to give the most important
correction to the semiclassical calculations discussed in the preceding
section. 

The derivation of the dephasing term of the weak localization correction
is described in Appendix B. The corresponding diagrams are shown in Fig.3. 
In general the result may be represented as a product of three
Cooperons, see Eqs.(\ref{b4}),(\ref{b6}). The whole process corresponds to 
the diffusion of an electron
from an initial point to some other point, where it emits a boson.
Then it travels to the second point and absorbs the same boson,
after that it goes back to the initial point. 
\begin{figure}
\centerline{\psfig{figure=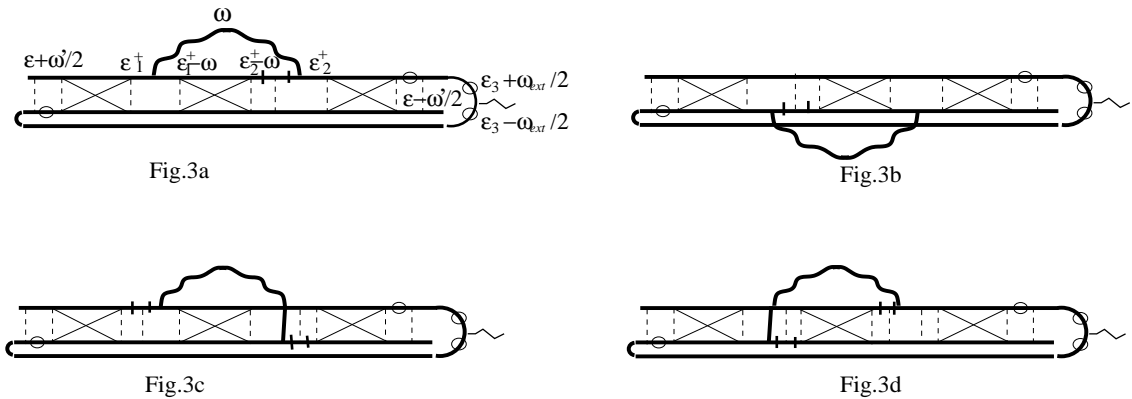,width=15 cm}}
\caption{
Four possible diagrams which contribute to the dephasing 
correction to the conductivity.
}
\end{figure}

In the case when the propagation of the emitted boson is described by the
Keldysh component of the boson Green's function, the Cooperons have the 
same time arguments at intermediate points so that the diffusive motion is
continuous in time, see Eq.(\ref{b4}).
Terms which have the retarded or advanced components of the boson Green's
function have a discontinuity at the points which represent the
nondiagonal vertices in the Keldysh space, Eq.(\ref{b6}). This 
discontinuity can be
explained by the fact that the energy of an electron is not conserved
along its trajectory because of the interaction with the classical field. 
We disregard these discontinuities and set $\tau=0$ in the Cooperons of  
Eq.(\ref{b6}). Then the integrals over $\tau$ are trivial. It is
possible to show that in this case we disregard terms which are of the
same order as higher order quantum corrections.

The dephasing correction to conductivity is
\be
\label{3.4}
\Delta\sigma_{deph}=\frac{4De^2}{\pi}
\int\frac{d\o}{2\pi}\int\frac{dq}{2\pi}\frac{2T}{\o}
f(\o/2T)J(\o,q)
\int_{0}^{+\infty}d \eta {\cal C}_2(T,\eta,\r),
\ee
where
\be
\label{3.5}
\begin{array}{c}
\displaystyle
{\cal C}_2(T,\eta,\r)=
2\int_{-\eta}^{+\eta}d
\zeta_1\int_{-\eta }^{\zeta_1}d \zeta_2\int d\r_1\int d \r_2
\left(\cos\o(\zeta_1-\zeta_2)-\cos\o(\zeta_1+\zeta_2)\right)
e^{-iq(\r_1-\r_2)}\times
\\
\displaystyle
{\cal C}(T,\eta,\zeta_1, \r,\r_1)
{\cal C}(T,\zeta_1,\zeta_2, \r_1,\r_2)
{\cal C}(T,\zeta_2,\eta, \r_2,\r).
\end{array}
\ee
with the thermal function 
\be
\label{3.6}
f(x)=\frac{x^2}{\sinh^2 x}-1.
\ee 
Note that $f(x)$ vanishes at $x\approx 0$, i.e in the infrared region,
because of the inclusion of classical fluctuations in the factors $\cal C$.
 
We can eliminate the integrals over $\r_{1,2}$ and represent ${\cal
C}_2(T,\eta,\r)$ as
\be
\label{3.7}
\begin{array}{c}
\displaystyle
{\cal C}_2(T,\eta,\r)=2 \int_{-\eta}^{+\eta}d
\zeta_1\int_{-\eta}^{\zeta_1}d \zeta_2
\left(\cos\o(\zeta_1-\zeta_2)-\cos\o(\zeta_1+\zeta_2)\right)
\\
\displaystyle
\exp\left(-i\q(\r(\zeta_1)-\r(\zeta_2))\right){\cal
C}(T,+\eta,-\eta,\r,\r),
\end{array}
\ee
and ${\cal C}(T,+\eta,-\eta,\r,\r)$ is given by Eq.(\ref{2.13}) without factor
$(2\pi\nu\t^2)^{-1}$. 

Now we can easily perform averaging over the fluctuations of the classical 
field according to Eq.(\ref{1.15}). As a result we find, that now we can 
substitute the Cooperon defined by Eq.(\ref{2.15}). The path integral can
be simplified by using the variables  
$R(t)$ and $x(t)$ introduced below Eq.(\ref{2.14}).   Since the
integrals over $R(t)$ and $x(t)$ are defined only at positive values of $t$, 
we have to rewrite Eq.(\ref{3.7}) as an integral over positive values of 
$\zeta_1$ and $\zeta_2$.  

In terms of the $R$ and $x$ variables the averaged ${\cal C}_2(T,\eta,\r)$ 
has the form
\be
\label{3.8}
\begin{array}{c}
\displaystyle 
{\cal C}_2(T,\eta,\r)=4\int_{0}^{+\eta}d
\zeta_1\int_{0}^{+\zeta_1}d \zeta_2 
\left(\cos\o(\zeta_1-\zeta_2)-\cos\o(\zeta_1+\zeta_2)\right)
\\
\displaystyle
\frac{1}{\sqrt{2}}
\int_{-\infty}^{+\infty}d R_1\int_{R(0)=R_1}^{R(\eta)=r/\sqrt{2}}
{\cal D}R(t)\exp\left(-\int_0^\eta dt \frac{\dot R^2(t)}{4D}\right)
e^{-iq(R(\zeta_1)-R(\zeta_2))/\sqrt{2}} \\
\displaystyle
\int_{x(0)=0}^{x(\eta)=0}
{\cal D}x(t)\exp\left(-\int_0^\eta dt 
\frac{\dot x^2(t)+\Omega^2 x^2(t)}{4D}\right)
e^{iqx(\zeta_2)/\sqrt{2}}(-2i)\sin\frac{q}{\sqrt{2}}x(\zeta_1). 
\end{array}
\ee

The integral over $R(t)$, written in the second line of Eq.(\ref{3.8}), 
can be done:
\be
\label{3.9}
\int_{-\infty}^{+\infty}d R_1\int_{R(0)=R_1}^{R(\eta)=r/\sqrt{2}}
{\cal D}R(t)\exp\left(-\int_0^\eta dt \frac{\dot R^2(t)}{4D}\right)
e^{-iq(R(\zeta_1)-R(\zeta_2))/\sqrt{2}}=e^{-D q^2(\zeta_1-\zeta_2)/2}.
\ee

The integral over the odd part of trajectories is more complicated. We will
represent the third line in the form:
\be
\label{3.10}
\begin{array}{c}
\displaystyle
\int_{x(0)=0}^{x(\eta)=0}
{\cal D}x(t)\exp\left(-\int_0^\eta dt 
\frac{\dot x^2(t)+\Omega^2 x^2(t)}{4D}\right)
e^{iqx(\zeta_2)/\sqrt{2}}\sin\frac{q}{\sqrt{2}}x(\zeta_1)=\\
\displaystyle
\int_{-\infty}^{+\infty}dx_1\int_{-\infty}^{+\infty}dx_2
\tilde {\cal C}(0,x_2,\zeta_2)\tilde {\cal C}(x_2,x_1,\zeta_1-\zeta_2)
\tilde {\cal
C}(x_1,0,\eta-\zeta_1)e^{iqx_2/\sqrt{2}}\sin\frac{q}{\sqrt{2}}x_1,
\end{array}
\ee
where 
\be
\label{3.11}
\begin{array}{c}
\displaystyle
\tilde {\cal C}(x_1,x_2,\zeta)=\int_{x(0)=x_2}^{x(\zeta)=x_1}
{\cal D}x(t)\exp\left(-\int_0^\zeta dt 
\frac{\dot x^2(t)+\Omega^2 x^2(t)}{4D}\right)= 
\\
\displaystyle
\sqrt{\frac{\Omega}{4\pi D \sinh \Omega\zeta}}
\exp\left(- \frac{\Omega}{4 D \sinh\Omega\zeta}\left( (x_1^2+x_2^2)\cosh
\Omega\zeta-2x_1 x_2\right) \right)
\end{array}
\ee
is the Green's function of a harmonic oscillator in imaginary time.

The integrals over $x_1$ and $x_2$ are Gaussian and can be done exactly for 
arbitrary value of $q$. In our case $q$ is bounded by $q_c\ll 1/L$.
It allows us to perform the expansion in powers of $q$. Keeping the 
first non--vanishing term we get
\be
\label{3.12}
\Delta\sigma_{deph}=\frac{-\Delta\sigma_{wl} \Omega^2}{I_0}
\int\frac{d\o}{2\pi}
f(\o/2T)
\int_{0}^{+\infty}d \eta I(\eta,\o),
\ee
where
\be
\label{3.13}
\begin{array}{c}
\displaystyle
I(\eta)=
\int_{0}^{+\eta}d
\zeta_1\int_{0}^{+\zeta_1}d \zeta_2
\left(\cos\o(\zeta_1-\zeta_2)-\cos\o(\zeta_1+\zeta_2)\right)
\frac{\sinh\Omega\zeta_2\sinh\Omega(\eta-\zeta_1)}{(\sinh
\Omega\eta)^{3/2}}=
\\
\displaystyle
\frac{1}{\Omega^2+\o^2}\left(\frac{\Omega\eta}{2\sqrt{\sinh\Omega\eta} }
+\frac{\Omega(\o^2-\Omega^2)}{4\o(\o^2+\Omega^2)}
\frac{\sin 2\o\eta}{\sqrt{\sinh\Omega\eta }}-
\frac{\Omega^2}{\o^2+\Omega^2}
\frac{\sin^2\o\eta \cosh \Omega\eta}{\sqrt{\sinh^3\Omega\eta}}
\right).
\end{array}
\ee

Now we can perform integral over frequency. The second and third terms in 
Eq.(\ref{3.13}) converge fast and we can use an approximation for the thermal 
function at small values: $f(x)=-x^2/3$. The first term can be evaluated 
in the lowest order in $\Omega/T$. As the result we get
\be
\label{3.14}
\Delta\sigma_{deph}=\Delta\sigma_{wl}\left( 1+ \frac{I_1}{4}\frac{\Omega}{T}
+\frac{2 I_2-I_3}{96}\frac{\Omega^2}{T^2}\right),
\ee
where $I_1$ was defined by Eq.(\ref{2.23}), 
\be
\label{3.15}
I_2=\int_0^\infty\frac{(1-2z)e^{-z}}{\sqrt{\sinh z}}dz\approx 1.02,
\ee
and
\be
\label{3.16}
I_3=\int_0^\infty\frac{(2z-1)e^{-2z}+1}{\sqrt{\sinh z}}\coth z dz
\approx 7.54.
\ee
The third term in the brackets is of second order in $\Omega/T$ and can be 
omitted.

We see that the quantum corrections are not important, unless $T\leq T_q$, 
with $T_q$ being defined by Eq.(\ref{2.16a}). Let us evaluate dephasing rate 
$\Omega(T_q)$ at $T=T_q$. According to Eq.(\ref{2.16a}) we have
\be
\label{3.17}
\Omega(T_q)=\frac{4DR_{eff}}{L^2}. 
\ee 
Here we use dimensionless resistance in terms of $\hbar/e^2$. 
We substitute $\Omega(T_q)$ to Eq.(\ref{2.16c}), see also Eq.(\ref{1.23}), 
and get
\be
\label{3.18}
\Delta\sigma_{wl}\sim -\sigma\sqrt{\frac{(R_w+R_0)^2}{R_0}}.
\ee
The weak localization becomes strong localization at $T\gg T_q$, 
if $(R_w+R_0)^2/R_0\gg 1$. In the opposite case $(R_w+R_0)^2/R_0\ll 1$ we have
$R_{eff} \ll 1$, consequently 
$\Omega(T_q)\ll D/L^2$ and the wire becomes zero--dimensional at high 
temperature. 
We conclude that in both limits the weak localization correction to 
conductivity deviates from the result of \cite{AAK} ( see Eq.(\ref{2.18}) 
at higher temperature than $T_q$ due to other reasons rather than due to 
the quantum corrections.

The plus sign before the second term in Eq.(\ref{3.14}) means that the 
dephasing rate found in \cite{AAK} is 
overestimated, and the quantum correction suppresses it.

\section{Electron--electron interaction}

Now we discuss how the result obtained in the previous section can be
generalized to the case of the real electron--electron interaction.
The screened Coulomb spectral
function of a dirty metal is given by Eq.(\ref{1.21}), where $\sigma_1$ is
the one dimensional conductivity of the wire and the momentum integral in
Eq.(\ref{1.20}) runs from $-\infty$ to $+\infty$. \cite{KA} We will use
the effective action in the form of Eq.(\ref{1.9}) with the appropriate
choice of mode density Eq.(\ref{1.21}) to make a semi-quantitative
calculation of the quantum
correction to the semiclassical result Eq.(\ref{4.5}).

First consider weak localization neglecting
the quantum correction.  This problem was solved in \cite{AAK}.  [See
also section 3 in the present paper.]  We discuss how this
result can be obtained by qualitative arguments.

The weak localization correction to conductivity is still given by 
Eq.(\ref{2.10}) 
and the Cooperon can be represented in the form of Eq.(\ref{2.13}).
Although Eq.(\ref{2.13}) was derived for small momentum $\q$ of the 
electric field
$V(\o,\q)$---see Appendix A---we can use it for the Coulomb
interaction with unbounded spectrum in the 
momentum space, since the main contribution comes from the 
long wavelength part of the interaction. In higher dimensions $D=2,\ 3$, the 
momentum integration has to be cut to satisfy conditions at which the 
Cooperon has been derived.

The next step is to average over the fluctuations of the classical 
field. The Cooperon is defined by Eq.(\ref{2.15}) with the `potential' $U(\r)$ 
given by the second line in Eq.(\ref{2.15a}).
Now the exponent is not an analytic function of the coordinate
variables $x(t)$.  

We can rewrite the exponent of Eq.(\ref{2.15}) in terms of dimensionless
variables
$y=r/L_{ee}$ (space) and $\zeta=t\gamma^{-1}$ (time). 
Then $L_{ee}$ and $\gamma^{-1}$ determine the characteristic 
space and time scales at which  the electron--electron interaction becomes 
important. At smaller scales 
the interaction is not important and the diffusion can be considered without
interaction. In the opposite case the diffusion process is 
suppressed by the interaction. According to Eq.(\ref{2.15}) and 
Eq.(\ref{2.15a}) $L_{ee}$ and $\gamma^{-1}$ are given by 
\be
\label{4.3}
\gamma=\frac{D}{L_{ee}^2}=\left(\frac{e^2T\sqrt{D}}{\sigma_1} 
\right)^{2/3}.
\ee
This result was found for the first time in \cite{AAK}.

This intrinsic cutoff may be introduced into the expression for
the weak localization correction for noninteracting electrons in the time 
representation, by cutting off the time integral with the exponential
function $\exp(-2t\gamma)$.
\be
\label{4.4}
\Delta \sigma_{wl}^{ee}\approx-\frac{4e^2D}{\pi}
\int_0^{\infty}e^{-2\gamma \tau}\frac{d\tau}{\sqrt{8\pi D
\tau}}=\frac{e^2}{\pi}\sqrt{\frac{D}{\gamma}}.
\ee
This expression is in agreement with the exact result found in \cite{AAK}.
[See Eq.(\ref{2.25}).]

The analysis of Appendix B can be directly applied to the 
case of electron--electron interaction with unbounded momentum integration. 
It follows that for the dephasing term of the weak localization correction to 
the conductivity we get Eq.(\ref{3.4}) with ${\cal C}_2(T,\eta,\r)$ given by 
Eq.(\ref{3.5}), which again can be represented in the form of Eq.(\ref{3.7}). 
Averaging over the fluctuations of the electric field $V(t,\r)$ gives an 
expression for ${\cal C}_2(T,\eta,\r)$, similar to that in Eq.(\ref{3.8}):
\be
\label{4.7}
\begin{array}{c}
\displaystyle 
{\cal C}_2(T,\eta,\r)=\int_{0}^{+\eta}d
\zeta_1\int_{0}^{+\zeta_1}d \zeta_2 
\left(\cos\o(\zeta_1-\zeta_2)-\cos\o(\zeta_1+\zeta_2)\right)
\\
\displaystyle
\int_{-\infty}^{+\infty}d R_1\int_{R(0)=R_1}^{R(\eta)=r/\sqrt{2}}
{\cal D}R(t)\exp\left(-\int_0^\eta dt \frac{\dot R^2(t)}{4D}\right)
e^{-iq(R(\zeta_1)-R(\zeta_2))} \\
\displaystyle
\int_{x(0)=0}^{x(\eta)=0}
{\cal D}x(t)\exp\left(-\int_0^\eta dt 
\left(\frac{\dot x^2(t)}{4D}+e^2\frac{\sqrt{2}}{\sigma_1} T
|x(t)|\right)\right)
e^{iqx(\zeta_2)}(-2i)\sin\frac{q}{\sqrt{2}}x(\zeta_1). 
\end{array}
\ee
We have changed variables for the path integral, introducing even
and odd parts, as discussed in section 2. The integral over the even
part of trajectories is not disturbed by the interaction and the result of 
integration is given by Eq.(\ref{3.9}). The path integral over $x(t)$ can be 
rewritten in the form of Eq.(\ref{3.10}), where $\tilde {\cal C}(t,x_1,x_2)$
is a solution to:
\be
\label{4.8}
\left(\frac{\partial}{\partial t}-D\nabla^2_{x_1}+\frac{\sqrt{2}e^2}{\sigma_1}
T|x_1|\right)
\tilde {\cal C}(t,x_1,x_2)=\delta(t)\delta(x_1-x_2).
\ee
This equation can be rewritten in the dimensionless variables, if time and 
space coordinates are divided by $\gamma^{-1}$ and $L_{ee}$. After 
this the solution can be found numerically. 

The solution to Eq.(\ref{4.8}) has the following important property:
at small time $t\ll \gamma^{-1}$ it is very similar to the Cooperon without 
interaction, but at time scale greater than $T_{ee}$ it is suppressed.  
To evaluate the right hand side of Eq.(\ref{3.12}), we can use the Cooperon 
without interaction, see Eq.(\ref{a3}), but introduce the upper 
cutoff for time integrals at $\gamma^{-1}$.
More exactly, we introduce exponent weight factor, which vanishes at time 
greater than $T_{ee}$, and substitute 
$\tilde {\cal C}(t,x_1,x_2)$ by the expression given by  
\be
\label{5.1a}
C(\o,k)=\frac{1}{i\o+Dk^2+\gamma}.
\ee 
In 
this case calculations can be easily completed. The dephasing part of the 
weak localization correction to conductivity is given by an analogue to
Eq.(\ref{3.12}):
\be
\label{4.8a}
\Delta\sigma_{deph}^{ee}=\frac{8De^4 T}{\pi\sigma_1}\int\frac{d\o}{2\pi}
f(\o/2T)\int_0^{\infty} d\eta I(\eta,\o),
\ee 
where
\be
\label{4.9}
\begin{array}{c}
\displaystyle
I(\eta,\o)=
\int_{0}^{+\eta} d x\int_{-\eta+x}^{+\eta-x} d y
\left(\cos\o x-\cos\o y\right)
\frac{\sqrt{x(\eta-x)}}{\eta}\exp(-2\eta\gamma).
\end{array}
\ee
The factor $e^{-\eta\gamma}$ 
produces the upper cutoff of the time integral.  
An analytical expression for $\int_0^{\infty}I(\eta)d\eta $ can be found.
This expression is cumbersome and we present only a term which
has the weakest frequency dependence. We would like to remind, that now we
consider high frequency contribution $\o\sim T $, since the low frequency
part is taken into account in terms of the classical field.
\be
\label{4.9a}
\int_0^{\infty}I(\eta,\o)d\eta = -\frac{\pi}{2}
\frac{\cos\left(\frac{3}{2}\arctan \frac{\o}{\gamma}\right)}
{\sqrt{\gamma^3}(\o^2+\gamma^2)^{3/4}}+\dots\sim
\frac{\pi}{(2\gamma\o)^{3/2}} .
\ee

Substituting this expression into Eq.(\ref{3.12}), we find 
an analogue of Eq.(\ref{3.14}):
\be
\label{4.10}
\Delta\sigma_{deph}^{ee}=J
\frac{e^2}{\pi}\sqrt{\frac{D}{T}}
\sim
\Delta\sigma^{ee}_{wl} 
\sqrt{\frac{\gamma}{T}},
\ee
where 
\be
\label{4.10a}
J=2 \int dx \frac{1}{x^{3/2}}\left(\frac{x^2}{\sinh^2 x}-1\right)\approx 
-3.66. 
\ee

Let us introduce the dimensionless conductance of the wire at length $L$:
\be
\label{4.11}
g(L)=\frac{\hbar}{e^2}\frac{\sigma_1}{L},
\ee
where $\sigma_1$ is the Drude (bare) conductivity of the wire.
Then Eq.(\ref{4.3}) can be rewritten in the form:
\be
\label{4.12}
\gamma=\frac{T}{g(L_{ee})}.
\ee
Since in the weak localization regime $g(L_{ee})\gg 1$, Eqs.(\ref{4.10}) and
(\ref{4.11}) mean that the quantum correction 
to the classical result obtained in \cite{AAK} is proportional to a small
quantity $[g(L_{ee})]^{-1/2}$. At sufficiently low temperature 
$g(L_{ee})$ approaches unity, but in this case the classical correction 
to the conductivity becomes comparable with the Drude conductivity $\sigma_1$, 
and the perturbation treatment of localization breaks down.  The conclusions
of this section and the last are thus the same:  in the region where the
correction to the conductivity may be described by the interference of two
time reversed trajectories, neglecting more complicated interference terms,
the quantum fluctuations of the dissipative environment are unimportant.  

\section{Discussion}

In this paper, we have constructed a bridge between semiclassical 
calculations \cite{AAK} which keep the interaction to all orders of 
perturbation theory and exact quantum mechanical calculations to first 
order in the interaction.\cite{AAG} 
While the semiclassical calculations are self 
consistent and do not require some external cutoff, the finite order 
perturbation theory is infrared divergent, and requires 
a low frequency cutoff. By contrast, our calculation of quantum corrections
is intrinsically regularized at low frequencies and the parameter of the 
perturbation theory 
is $\gamma/T$, see Eq.(\ref{4.10}).

These differences notwithstanding, we shall now show that
when the cutoff introduced in \cite{AAG} is treated as a parameter to be
determined self consistently, one obtains agreement with our
results and conclusions as presented in the last section.

The weak localization correction to conductivity found in
\cite{AAG}, Eq.(4.13a):
\be
\label{5.4a}
\Delta\sigma_{wl}=\Delta \sigma^{(0)}_{wl}+\Delta \sigma_{deph}+
\Delta \sigma'_{deph}+\Delta \sigma_{cwl}.
\ee  

The first term $\Delta\sigma_{wl}^{(0)}$ 
comes from the maximally crossed diagram without electron--electron
interaction with finite dephasing rate $\gamma$.
The result for $\Delta\sigma_{wl}^{(0)}$ is:
\be
\label{5.1}
\Delta\sigma_{wl}^{(0)}=-\frac{2De^2}{\pi}\int\frac{dk}{2\pi}C(\o=0,k)
=-\frac{e^2}{\pi}\sqrt{D\gamma^{-1}},
\ee
where $C(\o,k)$ is given by Eq.(\ref{5.1a}).

The first order correction due to the interaction is given by an equation
similar to 
Eq.(\ref{3.4}), see \cite{AAG}
\be
\label{5.2}
\Delta \sigma_{deph}=\frac{4De^2}{\pi}\int\frac{dk}{2\pi}
\int\frac{d\o}{2\pi}\int\frac{dq}{2\pi}\frac{2T}{\o}
F(\o/2T)J(\o,q) C_2(\o,\k,\q),
\ee
where
\be
\label{5.3}
\begin{array}{c}
\displaystyle
C_2(\o,\k,\q)=
\left(C^2(0,k)[C(-\o,k-q)+C(\o,k-q)]-
\right. \\
\displaystyle
\left.
C(-\o,k-q)C(\o,k-q)[C(0,k)+C(0,k-2q)]\right).
\end{array}
\ee
Here $F(x)=x^2/\sinh^2 x$ is the thermal function, which differs from 
$f(x)$ defined by Eq.(\ref{3.6}) by unity. [Recall that the unity in
Eq.(\ref{3.6}) eliminated the classical component in Eq.(\ref{3.4}), which
we treated to all orders of perturbation theory.] 

With the spectral function of the Coulomb interaction $J(\o, q)$  given
by Eq.(\ref{1.21}),
the integrals in Eq.(\ref{5.4}) can be done analytically in the limit 
$\gamma\ll T$, since the integral over $\o$ converges at $\o\sim\gamma$ and 
one can use the approximation $F(x)\approx 1$.
The result is \cite{AAG}:
\be
\label{5.4}
\Delta \sigma_{deph}=\frac{e^4DT}{4\pi\gamma^2\sigma_1}.
\ee
Assuming that the electron--electron interaction is the only
mechanism of decoherence, let us determine $\gamma$ 
self--consistently. Minimizing the sum
$\Delta \sigma^{(0)}_{wl}+\Delta \sigma_{deph}$ with respect to 
$\gamma$, we get $\gamma$ in agreement with Eq.(\ref{2.24}). 

The third term in Eq.(\ref{5.4a}) corresponds to the high frequency
contribution to the
dephasing term Eq.(\ref{5.2}) beyond the approximation $F(x)=1$ and has
the form:
\be
\label{5.4b}
\Delta \sigma'_{deph}=-\zeta\left(\frac{1}{2}\right)
\frac{e^4DT}{\pi\gamma^2\sigma_1}\sqrt{\frac{2\gamma}{\pi T }}.
\ee
The last term is the interaction correction to weak localization and it was 
presented in Eq.(\ref{b2}).

Let us substitute $\gamma$ from Eq.(\ref{2.24}) into Eq.(\ref{5.4a}).
The first two terms now become of the same order of magnitude and they
correspond
to the result of \cite{AAK}. The third term becomes
\be
\label{5.4c}
\Delta \sigma'_{deph}=4\zeta\left(\frac{1}{2}\right)
\Delta \sigma^{(0)}_{wl}\sqrt{\frac{2 \gamma}{\pi T}}\sim
\sqrt{\frac{1}{g(L_{ee})}}\Delta \sigma^{(0)}_{wl},
\ee
in agreement with our Eqs.(\ref{4.10},\ref{4.12}). 

The interaction correction of ref. \cite{AAG}, which we have not considered

\be
\label{5.4d}
\Delta\sigma_{cwl}=-3\zeta(3)\frac{e^2\sqrt{D\tau_H}}{2\pi\hbar}
\frac{e^2}{2\pi\hbar\sigma_1}\sqrt{\frac{\hbar D}{2\pi T}}\sim 
\sqrt{\frac{1}{g^3(L_{ee})}}\Delta \sigma^{(0)}_{wl}.
\ee
The last equality comes from 
\be
\label{5.4e}
\frac{e^2}{\hbar \s1}\sqrt{\frac{D}{T}}=
\frac{e^2\sqrt{D\gamma^{-1}}}{\hbar \s1}\sqrt{\frac{\gamma}{T}}\sim
\sqrt{\frac{1}{g^3(L_{ee})}}. 
\ee
Note that this correction is even smaller than $\Delta \sigma^{(1)}_{deph}$.

The reason for the smallness of these corrections is worth emphasizing:
the contribution of electromagnetic modes with frequencies greater than the
temperature is exponentially suppressed because detailed balance 
requires that dephasing be produced by the available electronic excitations
which have energies smaller than the temperature of the system.

At sufficiently low temperature the semiclassical result of \cite{AAK} 
breaks down, because the phase breaking length $L_\varphi$ becomes large
and the dimensionless conductance on that length scale $g(L_\varphi )\sim 1$. 
In this limit the weak localization picture for a disordered electron 
system with Coulomb interactions is not applicable and the notion of 
the dephasing rate becomes irrelevant to the problem of the quantum 
correction to the conductivity, which is no longer determined by 
the interference of two time reversed paths. 

In summary, we have in this paper dealt with interactions as a mechanism
for electronic dephasing by explicitly separating the low and high
frequency components.  The low frequency, or classical part, was shown to
reproduce the previously obtained results \cite{AAK}.
Then we treated the quantum component perturbatively and showed that it
contributes negligibly to the conductivity in the regime of weak
localization.  Finally, we showed that a recent cutoff dependent calculation
of quantum corrections \cite{AAG}, originally intended to apply only in
the presence of extrinsic phase-breaking effects, is in agreement with ours
if the cutoff is interpreted self consistently.

We conclude, that the electron--electron interaction via Coulomb force 
cannot explain the saturation of the dephasing rate in the weak 
localization regime. We believe that another mechanism is needed to explain 
the results of the experimental paper \cite{MJW}.

\section*{Acknowledgements}
The primary support for this research came from the Cornell Center for
Materials Research, funded by the National Science Foundation
under grant DMR-9632275.  The work was done at Cornell and several
other institutions during
the course of 1998.  VA thanks the Warden and Fellows of All Souls College,
Oxford for gracious hospitality from January to June; VA and MV thank the 
organizers of the 1998 Extended Summer Workshop at the ICTP, Trieste, for 
the opportunity to participate in this event during August, where they had
fruitful discussions with I. Aleiner and B. Altshuler.  We also thank
the ${\rm \O}$rsted Laboratory of the Niels Bohr Institute in the 
University 
of Copenhagen, and NORDITA for a friendly work environment from September
to December.  We have benefited from remarks and communications about ref
\cite{VA1} from N. Argaman, A. Kamenev, K. Matveev, and M. Reyzer, and 
also thank 
P. Hedegard and D. Khveshenko for stimulating remarks.

\appendix
\section{}
We discuss the derivation of the Cooperon in the presence of classical
electric field $V(t,\r)$. It is convenient to work in space--time
representation. 
The diagram equation for the cooperon is shown in Fig.2. The corresponding 
analytical equation has the form:
\be
\label{a1}
\begin{array}{c}
\displaystyle
C(t_1^+,t_1^-,t_2^+,t_2^-,\r,\r')=C^{(0)}(t_1^+,t_1^-,t_2^+,t_2^-,\r,\r')+
\\
\displaystyle
\int\dots\int dt_3^+dt_4^+dt_3^-dt_4^-d\r_3d\r_4
C^{(0)}(t_1^+,t_1^-,t_3^+,t_3^-,\r,\r_3)
\Sigma^+(t_3^+,t_3^-,t_4^+,t_4^-,\r_3,\r_4)
C(t_4^+,t_4^-,t_2^+,t_2^-,\r_4,\r')+
\\
\displaystyle
\int\dots\int dt_3^+dt_4^+dt_3^-dt_4^-d\r_3d\r_4
C^{(0)}(t_1^+,t_1^-,t_3^+,t_3^-,\r,\r_3)
\Sigma^-(t_3^+,t_3^-,t_4^+,t_4^-,\r_3,\r_4)
C(t_4^+,t_4^-,t_2^+,t_2^-,\r_4,\r').
\end{array}
\ee
We introduced notations:
\be
\label{a2}
\begin{array}{c}
\displaystyle
\Sigma^+(t_3^+,t_4^+,t_3^-,t_4^-,\r_3,\r_4)=
\int dt_5\int d\r_5 \R(t_3^+-t_5,\r_3-\r_5)
V(t_5,\r_5)
\\
\displaystyle
\R(t_5-t_4^+,\r_5-\r_4)\A(t_3^--t_4^-,\r_3-\r_4)
\\
\displaystyle
\Sigma^-(t_3^+,t_4^+,t_3^-,t_4^-,\r_3,\r_4)=
\int dt_5 \int d\r_5 \R(t_3^+-t_4^+,\r_3-\r_4)
\A(t_3^--t_5,\r_3-\r_5)
\\
\displaystyle 
V(t_5,\r_5)\A(t_5^--t_4^-,\r_5-\r_4),
\end{array}
\ee

$C^{(0)}(t_1^+,t_1^-,t_2^+,t_2^-,\r,\r')$ is the Cooperon without 
interaction. In the Fourier representation it is given by a ladder shown in 
Fig.2:
\be
\label{a3}
C^{(0)}(\o,\k)=\frac{1}{2\pi\nu\t}\frac{1}{1-(2\pi\nu\t)^{-1}
\Pi_0(\o,\k)}=\frac{1}{2\pi\nu\t}\frac{1}{D\k^2-i\o},
\ee
where 
\be
\label{a4}
\Pi_0(\o,\k)=\int\frac{d\p}{(2\pi)^d}\R(\e+\o,\p+\k)\A(\e,\p)=2\pi\nu\t
(1+i\o\t-D\k^2\t).
\ee

Using the definition of the Fourier transform of the Cooperon, see 
Eq.(\ref{2.11}), it is easy to prove the following relation:
\be
\label{a5}
C^{(0)}(t_1^+,t_1^-,t_2^+,t_2^-,\r,\r')=
\delta(t_1^++t_1^--t_2^+-t_2^-)
\int\frac{d\k}{(2\pi)^d}
\int\frac{d\o}{2\pi}e^{-\o(t_1^+-t_2^+-t_1^-+t_2^-)}e^{i\q(\r-\r')}
C^{(0)}(\o,\k).
\ee

Now consider the self energy of the Cooperon, which is the product of Green's 
functions given by Eq.(\ref{a2}).
In Eq.(\ref{a6}) we use equations
\be
\label{b3}
\begin{array}{c}
\displaystyle
\int\frac{d\p}{(2\pi)^d}\R(\e,p)\R(\e+\o,\p+\q)\A(\e',-\p+\k)\approx 2\pi i 
\nu\t^2 \\
\displaystyle
\int\frac{d\p}{(2\pi)^d}\A(\e,p)\A(\e+\o,\p+q)\R(\e',\k-\p)\approx -2\pi i 
\nu\t^2,
\end{array}
\ee
to get 
\be
\label{a6}
\begin{array}{c}
\displaystyle
\Sigma^{+}(t_3^+,t_3^-,t_4^+,t_4^-,\r_3,\r_4)=
2\pi\nu i\t^2 V(t_4^+,\r_4)\delta(t_3^+-t_4^+)\delta(t_3^--t_4^-)
\delta(\r_3-\r_4)\\
\displaystyle
\Sigma^{-}(t_3^+,t_3^-,t_4^+,t_4^-,\r_3,\r_4)=
-2\pi\nu i\t^2 V(t_4^+,\r_4)\delta(t_3^+-t_4^+)\delta(t_3^--t_4^-)
\delta(\r_3-\r_4).
\end{array}
\ee
Note, that Eq.(\ref{b3}) are obtained in the lowest order in $\o\t$ 
and $D\q^2\t$.

We see that the property of the Cooperon without interaction given by 
Eq.(\ref{a5}) allows us to find the interacting Cooperon in the form
\be
\label{a8}
C(t_1^+,t_1^-,t_2^+,t_2^-,\r,\r')=
{\cal C}(t_1^++t_1^-,t_1^+-t_1^-,t_2^+-t_2^-,\r,\r')
\delta(t_1^++t_1^--t_2^+-t_2^-).
\ee
Using Eqs.(\ref{a6}) and (\ref{a8}) we obtain the integral equation for the
Cooperon:
\be
\label{a7}
\begin{array}{c}
\displaystyle
{\cal C}(T,\eta_1,\eta_2,\r,\r')={\cal C}^{(0)}(T,\eta_1,\eta_2,\r,\r')+
\\
\displaystyle
2\pi\nu i \t^2\int\int d\eta_3 d\r_3
{\cal C}^{(0)}(T,\eta_1,\eta_3^+,\r,\r_3)
(V(T+\eta_3/2,\r_3)-V(T-\eta_3/2,\r_3))
{\cal C}(T,\eta_3,\eta_2,\r_3,\r').
\end{array}
\ee
This equation can be considered as a Dyson equation for the Cooperon in the
classical field.

Since the noninteracting cooperon satisfies
\be
\label{a9}
\left(\frac{\partial}{\partial \eta}- D\nabla^2\right)
{\cal C}^{(0)}(\eta,\r)=
\frac{1}{2\pi\nu\t^2}\delta(\eta)\delta(\r), 
\ee
we get the Shr\"odinger type equation for the interacting Cooperon:
\be
\label{a10}
\left(\frac{\partial}{\partial \eta}- D\nabla^2 -i\tilde V(T,\eta,\r,\r')
\right)
{\cal C}(T,\eta,\eta',\r)=
\frac{\delta(\eta-\eta')\delta(\r-\r')}{2\pi\nu\t^2}, 
\ee
where $\tilde V(T,\eta,\r)=V(T+\eta/2,\r)-V(T-\eta/2,\r)$.
Note that a constant electric field does not influence the Cooperon.

The solution to Eq.(\ref{a10}) can be represented in path integral form, given
by Eq.(\ref{2.13}).

\section{}

Now we will demonstrate how to get Eq.(\ref{3.4}) from Eq.(\ref{3.3}) for the 
impurity average value of the conductivity correction.

All diagrams which contribute
to the dephasing term ( see \cite{AAG} ) can be 
classified according to Fig.3 into four groups. 

First we consider diagrams, which contain the Keldysh component of 
the electric
field Green's function. Since it is coupled to electrons by two diagonal
in the Keldysh space vertices, the allowed positions for the  Keldysh  
component of the electron Green's function are situated near the current 
operator, as it is shown in Fig.3 by circles. Those diagrams are similar to
the diagrams which contribute to the leading part of the weak localization
correction to conductivity, see Appendix A.   

All four diagrams have the similat elements: the Hikami box, difined by 
Eq.(\ref{2.9}), three cooperons which are the solution of Eq.(\ref{a1}), and 
two Hikami boxes in the form of triangles, see Eq.(\ref{b3})

Note that the vertices do not depend on frequency, and substituting 
the time representation of the Cooperon, determined by Eq.(\ref{2.11}), we get
\be
\label{b4}
\begin{array}{c}
\displaystyle
\Delta\sigma_{K}=\frac{8e^2D}{\pi}\int d\r_1\int d\r_2\int d\eta
\int_{-\eta}^{+\eta} d\zeta_1\int_{-\eta}^{+\zeta_1}
{\cal L}^{(K)}(\o,\r_1-\r_2)
\times
\\
\displaystyle
{\cal C}(T,\eta,\zeta_1,\r,\r_1){\cal C}(T,\zeta_1,\zeta_2,\r_1,\r_2)
{\cal C}(T,\zeta_2,-\eta,\r_2,\r)\times \\
\displaystyle
\left(\cos \o(\zeta_2-\zeta_1)-\cos \o(\zeta_1+\zeta_2)
\right).
\end{array}
\ee
This expression contains the contribution in the lowest order in $\e_F \t$
to the dephasing correction to conductivity which has the Keldysh component 
${\cal L}^{(K)}(\o,\r)$ of the electric field Green's function.

Now we can consider the diagrams, which have the retarded or advanced 
components of the electric field Keldysh Green's function. The diagrams are
represented in Fig.3. In this case there are vertices which are not diagonal 
in the Keldysh space of electron Green's function.

We take as an example the diagram, of the first type.
The analytical expression has the following form:
\be
\label{b4a}
\begin{array}{c}
\displaystyle
\R(\e+\o'/2,\e_1^+,\r,\r_1)\K(\e_1^+-\o,\e_2^+,\r_1,\r_2)
\R(\e_2^+,\e_3+\o_{ext}/2,\r_2,\r')\\
\displaystyle
{\bf j}(\r'){\bf A}(\r')\A(\e_3-\o_{ext}/2,\e-\o'/2,\r',\r)
\frac{h(\e_3+\o_{ext}/2)-h(\e_3-\o_{ext}/2)}{\o_{ext}}
{\cal L}^{(R)}(\o,\r_1-\r_2).
\end{array}
\ee

The Keldysh component of electron Green's function is given by 
Eq.(\ref{1.19b}).
The dephasing correction to conductivity is given by diagrams 
with $\K_0$ situated at the positions shown in Fig.(3). Their sum is
\be
\label{b5}
\begin{array}{c}
\displaystyle
\int d\r_1\int d\r_2\int\frac{d\o}{2\pi}
\left({\cal L}^{(R)}(\o,\r_1-\r_2)-{\cal L}^{(A)}(\o,\r_1-\r_2)\right)
\R(\e+\o'/2,\e_1^+,\r,\r_1)\\
\displaystyle
\R(\e_1^+-\o,\e_2^+,\r_1,\r_2)
\R(\e_2^+,\e_3+\o_{ext}/2,\r_2,\r')\A(\e_3-\o_{ext}/2,\e-\o'/2,\r',\r)\\
\displaystyle
{\bf j}(\r'){\bf A}(\r')
h(\e_2-\o)
\frac{h(\e_3+\o_{ext}/2)-h(\e_3-\o_{ext}/2)}{\o_{ext}}.
\end{array}
\ee

Here we have added the term with ${\cal L}^{(A)}(\o,\r_1-\r_2)$ 
to the dephasing 
correction to conductivity. The same term has to be subtratcted from the 
interaction correction. This procedure was done in \cite{AAG}, where the 
interaction correction was calculated for the first time. The corresponding
analysis has to be done in the spirit of the present calculations taking the 
classical field in all oredrs of perturbation theory.

Now the diagram can be 
represented as a combination of Cooperons, defined by Eq.(\ref{2.11}), Hikami 
boxes, see Eq.(\ref{2.9}) and Eq.(\ref{b3}). Unlike the terms, which have 
the Keldysh propagator of the electric field, one vertex depends on 
electron energy, scattered by the electric field. In this case we
cannot perform integration over that energy, and we get a non--instananeous 
vertex in terms of the Cooperon coupling to the electric field. 
We get:
\be
\label{b6}
\begin{array}{c}
\displaystyle
\Delta\sigma_{I}=\frac{4e^2D}{\pi}\int d\r_1\int d\r_2\int d\eta
\int_{-\eta}^{+\eta} d\zeta_1\int_{-\eta}^{+\zeta_1}\int_{-\infty}^{+\infty} 
d\tau ({\cal L}^{(R)}(\o,\r_1-\r_2)-{\cal L}^{(A)}(\o,\r_1-\r_2))\times
\\
\displaystyle
{\cal C}(T,\eta,\zeta_1,\r,\r_1)
{\cal C}(T,\zeta_1,\zeta_2,\r_1,\r_2)
{\cal C}(T+\tau/2,\zeta_2+\tau,\tau-\eta,\r_2,\r)\times \\
\displaystyle
\exp(i\o(\zeta_2-\zeta_1))\int\frac{d\e_2}{2\pi}\int\frac{d\e_3}{2\pi}
e^{i(\e_2^+-\e_3)\tau}h(\e_2^+-\o)
\frac{h(\e_3+\o_{ext}/2)-h(\e_3-\o_{ext}/2)}{\o_{ext}}.

\end{array}
\ee
The conductivity of the system is determined by the average value with 
respect to the fluctuations of the electric field, see Eq.(\ref{1.19}), 
we consider the average 
value of the expression in the second line of Eq.(\ref{b6}).
Substituting the Cooperons in the form of Eq.(\ref{2.13}), we get for the 
interaction term of the cooperon action, see the second term in 
Eq.(\ref{2.15}):
\be
\label{b7}
\begin{array}{c}
\displaystyle
S_{int}[\tau,\r(t)]=\frac{2e^2T}{\sigma_1}
\int\frac{d\q}{(2\pi)^d}\frac{1}{q^2}e^{i\q(\r(t_1)-\r(t_2))}\\
\displaystyle
\left(
\int_{-\eta}^{\zeta_2}dt_1\int_{-\eta}^{\zeta_2}dt_2
(\delta(t_1-t_2)-\delta(t_1+t_2))+\right. \\
\displaystyle
\int_{\zeta_2}^{+\eta}dt_1\int_{\zeta_2}^{+\eta}dt_2
(\delta(t_1-t_2)-\delta(t_1+t_2))+\\
\displaystyle
\left.
\int_{-\eta}^{\zeta_2}dt_1\int_{\zeta_2}^{\eta}dt_2
(\delta(t_1+\tau-t_2)+\delta(t_1-\tau-t_2)-\delta(t_1+t_2+\tau)-
\delta(t_1+t_2-\tau))\right) \\
\end{array}
\ee
We replace $S_{int}[\tau,\r(t)]$ by $S_{int}[\tau=0,\r(t)]$ and consider the
difference $\Delta^{I}_{int}=
S_{int}[\tau,\r(t)]-S_{int}[\tau=0,\r(t)]$ as perturbation.
The first order term in $\Delta^{I}_{int}$ is of the same order as the 
higher order terms in the quantum part of the influence functional and can be 
neglected for our purposes.

So we can neglect the dependence on $\tau$ of the second line in 
Eq.(\ref{b6}), and perform integration over $\tau$. After that integrals 
over energies $\e_2$ and $\e_3$ can be done.

The remaining three diagrams can be similarly calculated. Collecting 
all terms together, see also Eq.(\ref{b4})
we get Eq.(\ref{3.4}).  

\section{}

In this appendix we discuss a recent paper by \cite{GZ} Golubev and Zaikin,
which obtains a saturating dephasing time in the limit of zero temperature,
a results very different from \cite{AAK} and \cite{AAG}.  In our opinion the 
difference is due to uncontrolled approximations.  
It is important to note that the contradiction between \cite{GZ} and \cite{AAK}
appears for $g(L_\varphi)\gg 1$, in which region the corrections to \cite{AAK}
have been shown in the present paper to be negligible.

Before discussing details, some general remarks are in order.   The result
of \cite{GZ} is infrared divergent, see Eq.(76) in \cite{GZ}.  
In the present paper, we have shown that a consistent
calculation does not require an extrinsic cut-off.  This suggests that by
dropping terms the authors of \cite{GZ} have lost some essential physics.
We would also like to mention that the authors of \cite{GZ}, 
have referred to the Caldeira--Leggett model
of a quantum particle coupled to environment \cite{CL} as a similar model, 
which has to be treated in the same nonperturbative manner as paper
\cite{GZ} claims 
to treat the weak localization. However, in the solution of the
Caldeira--Leggett model, physical quantities are free
of unphysical divergences. For example, Eq.(83) in \cite{A?} for the 
average value of the particle momentum is expressed as an integral with
an intrinsic infrared cutoff. 

In what follows we shall take some formulas from \cite {GZ} on faith, and
question subsequent approximations.  The following expression is used 
to calculate the conductivity of the electron system:
\be
\label{5.11}
\sigma=-\frac{2e^2}{3m}\int_{-\infty}^{0}dt' \int\frac{d\p}{(2\pi)^d}
\hat W(t')\p\frac{\partial n(\p)}{\partial p},
\ee
with $\partial n(\p)/\partial\p \approx -{\bf v}_F \delta(\e)$.
The evolution operator $\hat W(t')$ is
\be
\label{5.12}
\hat W(t')=U_l(t')\cdot U_r(t'), 
\ee
where
\be
\label{5.13}
\begin{array}{c}
\displaystyle
U_l(t)={\rm T}\exp\left(-i\int_0^t d\tau H_l(\tau,\r_1,\r_2)\right),\\
\displaystyle
U_r(t)={\rm T}\exp\left(i\int_0^t d\tau H_r(\tau,\r_1,\r_2)\right)\\
\end{array}
\ee
and the Hamiltonians in the interaction picture are:
\be
\label{5.14}
\begin{array}{c}
\displaystyle
H_l(t,\r_1,\r_2)=-eV^+(t,\r_1)\delta(\r_1-\r_2)-
\frac{1}{2}(\delta(\r_1-\r_2)-2\rho(\r_1,\r_2))eV^-(t,\r_2),
\\
\displaystyle
H_r(t,\r_1,\r_2)=-eV^+(t,\r_1)\delta(\r_1-\r_2)+
\frac{1}{2}(\delta(\r_1-\r_2)-2\rho(\r_1,\r_2))eV^-(t,\r_1).
\end{array}
\ee
Here $V^+(t,\r)$ and $V^-(t,\r)$ are the components of the fluctuating 
electric field.

The conductivity defined by Eq.(\ref{5.11}) has to be averaged over 
fluctuations of the electric field to reproduce the conductivity of the 
interacting electron system. As the result of this averaging the authors of
\cite{GZ} obtain the following effective action for the electrons:
\be
\label{5.11aa}
S[t,\r(t)]=S_R[t,\r]+iS_I[t,\r],
\ee
where
\be
\label{5.5}
\begin{array}{c}
\displaystyle
S_I[t,\r,\dot \r]=\frac{e^2}{2}\int_{-t}^{+t}dt_1\int_{-t}^{+t}dt_2
\left(2I(t_1-t_2,\r(t_1)-\r(t_2))\right.
\\
\displaystyle
\left.
-I(t_1+t_2,\r(t_1)-\r(t_2))-
I(-t_1-t_2,\r(t_1)-\r(t_2)) \right).
\end{array}
\ee 
\be
\label{5.8}
\begin{array}{c}
\displaystyle
S_R[\r_1,\p_1,\r_2,\p_2]=\frac{e^2}{2}\int_{-t}^{+t}dt_1\int_{-t}^{+t}dt_2
\left(R(t_1-t_2,\r_1(t_1)-\r_1(t_2))(1-2n(\p_1(t_2),\r_1(t_2)))-\right. \\
\displaystyle
R(t_1-t_2,\r_2(t_1)-\r_2(t_2))(1-2n(\p_2(t_2),\r_2(t_2)))+\\
\displaystyle
R(t_1-t_2,\r_1(t_1)-\r_2(t_2))(1-2n(\p_2(t_2),\r_2(t_2)))-\\
\displaystyle
\left.
R(t_1-t_2,\r_2(t_1)-\r_1(t_2))(1-2n(\p_1(t_2),\r_1(t_2)))\right).
\end{array}  
\ee
We have shifted the limits of time integration, so that the sum of the upper
and lower limits is equal to zero. $I(t,\r)$ and $R(t,\r)$  are given by
\be
\label{5.5a}
I(t,\r)=\int\frac{d\o d^d\k}{(2\pi)^{d+1}}{\rm
Im}\left(\frac{-4\pi}{k^2\varepsilon(\o,\k)}\right) \coth\frac{\o}{2T}e^{-i\o
t+i\k\r},
\ee
\be
\label{5.9}
\begin{array}{c}
\displaystyle
R(t,\r)=R_e(t,\r)+R_o(t,\r), \\
\displaystyle
R_e(t,\r)=\int\int\frac{d\o d\k}{(2\pi)^{d+1}}{\rm
Re}\frac{4\pi}{k^2\varepsilon(\o,\k)}\cos\o t e^{i\k\r} \\
\displaystyle
R_o(t,\r)=\int\int\frac{d\o d\k}{(2\pi)^{d+1}}{\rm
Im}\frac{4\pi}{k^2\varepsilon(\o,\k)}\sin\o t e^{i\k\r}.
\end{array}
\ee
Let us consider the conclusions that were drawn from these formulas.
The authors of \cite{GZ} omitted the second 
and third terms  on the right hand side of the expression for the imaginary
part of the action $S_I[t,\r(t)]$, Eq.(\ref{5.5}). 
Their argument for doing this is that the first term increases linearly in 
time $t$, while both terms in the second line have slower long time dependence.
We agree that the first term is indeed important 
in the limit of very large $t$. But the contribution to 
conductivity is determined by time $t=\tau_{\varphi}$ 
which satisfies $S_I(\tau_{\varphi})\sim 1$, calling the argument into
question. 

Indeed, one can see the correspondence between the action Eq.(\ref{5.5}) 
and the semiclassical calculations of \cite{AAK}, see also section III.
The form of Eq.(\ref{5.5a}) 
is similar to that found by \cite{AAK}. If we use
${\rm Im}\e^{-1}(\o,\k)=\o/4\pi\sigma$, ( see Eq.(74) in \cite{GZ} ), we get 
\be
\label{5.6}
S_I[t,\r(t)]\sim \int_{-t}^{+t}dt_1 
\int\frac{d\k}{(2\pi)^d}\frac{1}{\k^2}\left(1-e^{i\k(\r(t_1)-\r(-t_1))}
\right)
\ee
Here we have used the fact, that the integral over frequencies gives
a narrow function of $t_1-t_2$ which vanishes for $|t_1-t_2|\geq 1/T$. Then 
keeping in mind the inequality $T\tau_\varphi\gg 1$ we can approximate that 
function by $\delta(t_1-t_2)$.

The unity in the brackets corresponds to the term considered in \cite{GZ}.
[See their Eq.(71).] It originates from the first term in Eq.(\ref{5.5}):
The second term in Eq.(\ref{5.6}), which originates from the second line in 
Eq.(\ref{5.5}) was neglected in \cite{GZ}. Indeed, one
is allowed to neglect it when the exponent $\k(\r(t_1)-\r(t_2))$ is large. 
However, for the long wavelength ($|\k| \to 0 $) modes which enter the
integral in Eq.(\ref{5.6}), one has 
to keep this term.

The real part of the effective action contains the factor $(1-2n(\p,\r))$. 
The authors of \cite{GZ} neglected the time  
dependence of the momentum. We expect that the $\p-$dependence  of the 
action is crucial for obtaining the correct result. 
It is worth mentioning that $n(\p,\r)$ is a sharp function of momentum 
near the Fermi surface of the electrons. The approximation of 
a sharp function by its value at a particular point is at best dangerous.
We will present arguments which raise questions about its validity. 
For this purpose we rewrite 
the real part of the action $S_R$, Eq.(\ref{5.8}) in another form:
\be
\label{5.10}
\begin{array}{c}
\displaystyle
S_R[\r(t),\p(t)]=\frac{e^2}{2}\int_{-t}^{+t}dt_1\int_{-t}^{+t}dt_2
(1-2n(\p(t_2),\r(t_2))) \\
\displaystyle
\left(
R_o(t_1-t_2,\r(t_1)-\r(t_2))-R_o(t_1+t_2,\r(t_1)-\r(t_2)) \right).
\end{array}
\ee 
Eq.(\ref{5.10}) is derived for the time reversed paths
$\r_1(t)=\r_2(-t)=\r(t)$, 
that is for the same class of trajectories as Eq.(68) in \cite{GZ}. We 
represented it in the most transparent form, so that a complicated factor 
$(1-2n(\p(t_2),\r(t_2)))$ is taken at the same moment of time for all terms 
of the integrand in Eq.(\ref{5.10}). We see, that if $n(\p(t))$ 
is a constant or an even function of time $t$, the real part of the effective 
action vanishes. In the opposite case, when $n(\p(t))$ is odd, the real 
part of the effective action is of the same order, as the imaginary 
part, taken at $T=0$. 

Now let us understand if variations of $n(\p(t))$ may be important.

We can introduce the eigenstates of the system without interaction.
In the interaction picture we have states $|\e,j,t>$, where $\e$ is the 
energy of the state, $j$ labels the degenerate levels corresponding to 
$\e$. We have the following equality: 
\be
\label{5.15}
<\e,j,t|\e',j',t'>=\delta_{\e,\e'}
\delta_{jj'}e^{-i\e(t-t')}.
\ee
In the eigenstate basis the density matrix can be represented as
\be
\label{5.16}
\rho= |\e,j,t>\frac{1}{\exp(\e/T)+1}<\e,j,t|. 
\ee

Then, according to Eq.(\ref{5.11}, \ref{5.12}) we have to evaluate 
\be
\label{5.17}
M_{jj'}(\e,\e')=\sum_{|\e_i|\sim T,j_i}\overline{
<e_l,j_l,t|U_l(t)|\e_i,j_i,0><\e_i,j_i,0|U_r(t)|\e_r,j_r,t>},
\ee
here the sum over the initial states is extended over the energy range, 
comparable with the temperature of the system and $\overline{(\dots)}$ 
means average over the electric field fluctuations. 
 
The evolution operators, Eq.(5.13), are understood as an expansion of the 
exponential functions. Then the right hand side of Eq.{5.17} is a series of 
matrix elements, which we denote as $M^{(n,m)}_{jj'}(\e,\e')$, where $n,m$
shows the order of the expansion of $U_l(t)$ or $U_r(t)$. As an example 
let us take the third term
of the series for $U_l(t)$ and the first term of $U_r(t)$.
We have
\be
\label{5.18}
\begin{array}{c}
\displaystyle
M^{(2,0)}_{jj'}(\e,\e')=-\sum_{|\e_i| \sim T ,j_i}
\int_0^t dt_1 \int_0^{t_1} dt_2 
\\
\displaystyle
<\e,j,t|\overline{H_l(t_1)H_l(t_2)}|\e_i,j_i,0>
<\e_i,j_i,0|\e',j',t>=
\\
\displaystyle
\int\int\frac{d\o d\q}{(2\pi)^{d+1}}\sum_{\e_2,j_2}\sum_{|\e_i| \sim T ,j_i}
\int d\r_1\int d\r_2
\\
\displaystyle
\int_0^t d t_1\int_0^{t_1} d t_2 e^{i \e_r(t-t_1)}e^{i (\e_2+\o)(t_1-t_2)}
e^{i\e_i t_2}
\\
\displaystyle
<\e,j,t_2|e^{i\q\r_1}|\e_2,j_2,t_2><\e_2,j_2,t_1|e^{-i\q\r_2}|\e_i,j_i,t_1>
\delta_{\e_i,\e'}\delta_{jj'}
\\
\displaystyle
\left(\frac{1}{2}\left(\coth\frac{\o}{2T}+\tanh\frac{\e_2}{2T}\right)
{\rm Im}\frac{4\pi}{ q^2 \varepsilon(\o,\q)}+
i \tanh\frac{\e_2}{2T}{\rm Re} \frac{4\pi}{q^2\varepsilon(\o,\q)}\right),
\end{array}
\ee
where $\chi(\o,\q)$ is the susceptibility of the electron system.

The first term in the last line of Eq.(\ref{5.18}) comes from the 
$\overline{V^+(t)V^+(0)}$ correlation function and the other contain 
imaginary part of
$\overline{V^-(t)V^+(0)}$. $\overline{V^-(t)V^-(0)}$ is identically zero and 
$\overline{V^+(t)V^-(0)}$ does not contribute to the above matrix element 
due to causality.

In ref. \cite{GZ} only the imaginary part of the dielectric susceptibility 
of the electron system contributes to the weak localization correction.
That is why we consider only the part of the matrix element 
$M_{jj'}^{(2,0)}(\e,\e')$, which contains ${\rm Im}\varepsilon^{-1}(\o,\q)$.
[ It was shown in \cite{AAG}, that terms with 
${\rm Re}\varepsilon^{-1}(\o,\q)$ contribute to the interaction correction
to weak localization, and do not contribute to the dephasing correction 
to conductivity. ]
Since the characteristic values of the upper limits of time integrals in 
Eq.(\ref{5.18}) are of the order of dephasing time $\tau_{\phi}$, 
the integral over $t_2$ vanishes, unless $|\e_2+\o-\e_i|\tau_{\phi}< 1$.
We have $\tau_{\phi} T\gg 1$ and consider the high frequencies
$\o\gg T$. Since $\e_i\sim T $, we see that the contribution
of high frequencies is exponentially suppressed. 
One can repeat this analysis for other terms $M^{(n,m)}_{jj'}(\e,\e')$ 
and observe the same type of cancellation of high frequency contributions to
the conductivity in any order of perturbation theory. 
A similar discussion can be found in \cite{EH}.

Note, that one can take sum 
$$ 
M^{(2,0)}_{jj'}(\e,\e')+
M^{(1,1)}_{jj'}(\e,\e')+M^{(0,2)}_{jj'}(\e,\e').
$$
This expression reduces to the quantum correction found in \cite{AAG}, 
but in this case one has to perform a sum over the spectrum of the 
disordered electron system. The expression would be divergent at $\q\to 0$.
To cure this divergence one should sum all orders of perturbation 
theory or introduce an external magnetic field to remove the long wave 
divergence.

We emphasize again that the paper \cite{GZ}  
predicts a saturation in the dephasing rate when the inequality 
$\tau_{\phi} T\gg 1$ is still satisfied. Our discussion in this appendix is 
based exactly on the same inequality. And in this case we see that 
terms in the second line of Eq.(\ref{5.5}) have to be kept and there is no 
contribution from electromagnetic modes with excitation energy much greater 
than temperature.  In short, the `nonperturbative' calculations of \cite{GZ}
disregarded terms which we have shown to be important.

\end{document}